\documentclass[conference, letterpaper]{IEEEtran}

\IEEEoverridecommandlockouts

\usepackage{hyperref}
\usepackage{float}
\usepackage[caption=false,font=footnotesize,labelfont=sf,textfont=sf]{subfig}
\usepackage{booktabs} 
\usepackage{amsmath,amsthm,amssymb}
\usepackage{mathtools} 
\usepackage{pdfpages}
\usepackage{pdflscape}
\usepackage{graphicx}
\usepackage[linesnumbered,ruled]{algorithm2e}
\usepackage[capitalize]{cleveref}
\usepackage{changes}
\usepackage{balance}
\usepackage{soul}
\usepackage{multirow}
\usepackage{hhline}
\usepackage{diagbox}
\usepackage{enumitem}
\usepackage{makecell}
\usepackage{colortbl}
\usepackage{lettrine}
\usepackage[letterpaper, total={516.0pt,682.0pt}]{geometry} 
\usepackage{siunitx} 
\usepackage{multirow}
\usepackage[noadjust]{cite}

\usepackage{etoolbox} 

\makeatletter
\patchcmd{\algocf@makecaption@ruled}{\hsize}{\textwidth}{}{} 
\patchcmd{\@algocf@start}{-1.5em}{0em}{}{}
\patchcmd\algocf@Vline{\vrule}{\vrule \kern-0.4pt}{}{}
\patchcmd\algocf@Vsline{\vrule}{\vrule \kern-0.4pt}{}{}
\makeatother

\AtBeginEnvironment{algorithm}{\linespread{0.93}\selectfont}

\AtEndEnvironment{algorithm}{\linespread{1.0}\selectfont}

\begin{document}

\title{Optimal Multi-Constrained Workflow Scheduling for Cyber-Physical Systems in the\\ Edge-Cloud Continuum}

\author{\IEEEauthorblockN{Andreas Kouloumpris, Georgios L. Stavrinides, Maria K. Michael, and Theocharis Theocharides}
 \IEEEauthorblockA{\textit{KIOS Research and Innovation Center of Excellence} \\
 \textit{Department of Electrical and Computer Engineering, University of Cyprus}\\
 Nicosia, Cyprus \\
 Email: \{kouloumpris.andreas, stavrinides.georgios, mmichael, ttheocharides\}@ucy.ac.cy}
 }
 
\maketitle

\begin{abstract}
The emerging edge-hub-cloud paradigm has enabled the development of innovative latency-critical cyber-physical applications in the edge-cloud continuum. However, this paradigm poses multiple challenges due to the heterogeneity of the devices at the edge of the network, their limited computational, communication, and energy capacities, as well as their different sensing and actuating capabilities. To address these issues, we propose an optimal scheduling approach to minimize the overall latency of a workflow application in an edge-hub-cloud cyber-physical system. We consider multiple edge devices cooperating with a hub device and a cloud server. All devices feature heterogeneous multicore processors and various sensing, actuating, or other specialized capabilities. We present a comprehensive formulation based on continuous-time mixed integer linear programming, encapsulating multiple constraints often overlooked by existing approaches. We conduct a comparative experimental evaluation between our method and a well-established and effective scheduling heuristic, which we enhanced to consider the constraints of the specific problem. The results reveal that our technique outperforms the heuristic, achieving an average latency improvement of 13.54\% in a relevant real-world use case, under varied system configurations. In addition, the results demonstrate the scalability of our method under synthetic workflows of varying sizes, attaining a 33.03\% average latency decrease compared to the heuristic.
\end{abstract}


\section{Introduction}
\label{sec:intro}

The edge-hub-cloud paradigm has emerged as an alternative to the conventional edge computing concept.
This shift is driven by the increasing demand for intelligent decision-making and real-time response at the network edge \cite{Zheng2021, Alam2017, Kashino2019, Savva2021}.
In contrast to the traditional concept that typically has three distinct layers within the edge-cloud continuum (i.e., edge devices, edge servers, and cloud servers), this paradigm considers edge and hub devices that are usually battery-powered in the bottom layer, and cloud servers in the top layer. A hub device (e.g., a smartphone or laptop) often has a higher computational capacity than an edge device (e.g., a wearable or single-board computer). Furthermore, it is physically closer to the edge devices than an edge server (although less capable), facilitating their communication with the remote cloud data center.

With the advent of the edge-hub-cloud paradigm, novel cyber-physical applications have been developed, consisting of precedence-constrained tasks that require diverse device capabilities for their execution, such as specific sensors, actuators, or software/hardware modules \cite{Liu2022}. 
These applications, referred to as workflows, encompass a wide range of latency-critical scenarios. 
The use of biomedical edge devices with integrated sensors and actuators for remote patient monitoring and support (e.g., smart pacemakers and wearable cardioverter defibrillators) \cite{Zheng2021, Alam2017} or the use of unmanned aerial vehicles (UAVs) with different capabilities for autonomous critical infrastructure inspection or search-and-rescue missions (e.g., UAVs with thermal/multispectral cameras or customized payload release systems) \cite{Kashino2019, Savva2021}, are some prominent examples.
In these scenarios, multiple edge devices with distinct capabilities cooperate with a hub device and a cloud server, which may also feature specialized software or hardware components, like libraries or accelerators for machine learning inference.

The intrinsic criticality of such applications requires \emph{optimal} task scheduling to determine where to allocate and when to execute each task, so as to achieve the minimum possible latency, which is usually bounded by a strict deadline \cite{Tang2019}. 
In contrast to heuristic approaches, exact methods such as mixed integer linear programming (MILP) can yield optimal schedules, but are typically computationally intensive. 
However, the pre-programmed nature of the examined applications allows the use of these methods for offline scheduling---although solutions should still be provided in a reasonable time to be practical.
In exact methods, time is modeled as either discrete or continuous.
Discrete-time models simplify the problem by considering that events occur only at the beginning or end of predefined time intervals, which is not a realistic assumption.
On the other hand, continuous-time models offer higher accuracy by allowing events to occur at any time. However, this flexibility makes the problem formulation more challenging, particularly the modeling of time-dependent cumulative constraints for the concurrent utilization of limited resources by multiple tasks, such as main memory, storage, or specific device capabilities \cite{Floudas2005}.

Moreover, challenges arise due to the heterogeneity of the devices where the considered applications are deployed, as they often feature different multicore processors and varied sensing/actuating capabilities. 
These challenges are amplified by the diverse communication characteristics of the operating setting.
In addition, moving from the cloud towards the network edge, there are increasing limitations in the computational, memory, storage, and energy capacities of the devices. 
In such environments, it is imperative to employ a workflow scheduling strategy that takes into account all these constraints, which make finding an optimal schedule particularly challenging.
To this end, we propose an offline continuous-time MILP approach to optimally schedule a workflow application in an edge-hub-cloud cyber-physical system (CPS) with multiple edge devices, heterogeneous multicore processors, and distinct device capabilities. Our objective is to minimize the overall latency, while considering crucial constraints that characterize these systems.

The contributions of this work are summarized as follows:
\begin{itemize}
\item We propose a comprehensive continuous-time MILP formulation to optimally schedule a workflow application in the  examined edge-hub-cloud CPS by minimizing the overall latency. Our formulation is facilitated by an extended representation of the application task graph.

\item We holistically address multiple constraints that are overlooked by existing scheduling methods, both exact and heuristic. These constraints are based on the memory, storage, and energy limitations of the devices, the heterogeneity and multicore architecture of the processors, the diverse device capabilities, the execution deadline, as well as the computational and communication latency and energy requirements of the tasks.

\item As the proposed approach is the first to provide\,an\,optimal solution for the considered problem, constraints, and architecture, we evaluate and compare it against a widely used and effective scheduling heuristic, the heterogeneous earliest finish time (HEFT) \cite{Topcuoglu2002, Kuhbacher2019, Aldegheri2020}. To ensure a fair and meaningful comparison, we extended HEFT to incorporate the constraints of the examined problem.

\item In our experiments, we consider a real-world workflow for the autonomous UAV-based power infrastructure inspection, under various system configurations. 
To further validate and investigate the scalability of our method to applications of different sizes, we utilize multiple synthetic workflows with appropriate parameters.
\end{itemize}

The rest of the paper is organized as follows. \cref{sec:related} provides an overview of related literature. \cref{sec:framework} describes the proposed MILP approach, whereas \cref{sec:heft} explains our extension to HEFT. \cref{sec:evaluation} presents the experimental setup, the evaluation results, and the empirical scalability analysis of our technique. \cref{sec:conclusion} provides concluding remarks.

\section{Related Work}
\label{sec:related}
Workflow scheduling in distributed environments is a well-studied problem that has been investigated extensively using both exact and heuristic methods \cite{Genez2020, Mo2022, Mo2023, Wu2018, Zengen2020, Bai2021, Alsurdeh2021, Stavrinides2019, Topcuoglu2002, Kuhbacher2019, Aldegheri2020}.
For example, MILP-based scheduling techniques for deadline-constrained workflows are proposed in \cite{Genez2020, Mo2022, Mo2023, Wu2018, Zengen2020}, with \cite{Mo2023} and \cite{Zengen2020} focusing on specific CPS architectures.
While these methods \cite{Genez2020, Mo2022, Mo2023, Wu2018, Zengen2020} provide optimal schedules, none of them considers a multi-tier setting or the memory and storage limitations of the devices. 
The approaches in \cite{Mo2023} and \cite{Zengen2020} also assume single-core processors, whereas homogeneous (rather than heterogeneous) multicore processors are examined in \cite{Mo2022}.
Moreover, only \cite{Mo2022, Mo2023, Wu2018} address energy consumption.   
Even though the technique in \cite{Mo2023} considers different sensing/actuating capabilities among CPS devices, it is limited to only one sensor or actuator per device. 
On the other hand, no device capabilities are considered in \cite{Zengen2020}, although a CPS is examined.
Regarding the representation of time in the problem formulation, the methods in \cite{Genez2020} and \cite{Zengen2020} employ discrete-time MILP, in contrast to \cite{Mo2022, Mo2023, Wu2018} that use continuous-time MILP.

\begin{table}[t]
\setlength{\tabcolsep}{2pt}
\centering
\caption{Comparison With Related Research Efforts}
\vspace{-5pt}
\label{table:comparison}
\resizebox{\columnwidth}{!}{
    \begin{tabular}{lcccccccc} 
        \toprule
        \multirow{3}{*}{Reference} & \multicolumn{5}{c}{Considered Constraints} & Optimal & Multi-tier & Multicore\\
        \cline{2-6}
                                   & Deadline & Capability & Memory & Storage & Energy & Solution & Environ. & Processors\\
                                   & & (s/m)${}^{1}$ & & & & (d/c)${}^{2}$ & & (h/H)${}^{3}$\\
        \hline
        \cite{Topcuoglu2002} & - & - & - & - & - & - & - & \checkmark (H)\\
        \cite{Kuhbacher2019} & - & - & - & - & - & - & - & \checkmark (H)\\
        \cite{Aldegheri2020} & - & - & - & - & - & - & - & \checkmark (H)\\

        \cite{Genez2020} & \checkmark & - & - & - & - & \checkmark (d) & - & \checkmark (H)\\ 
        \cite{Mo2022} & \checkmark & - & - & - & \checkmark & \checkmark (c) & - & \checkmark (h)\\ 
        \cite{Mo2023} & \checkmark & \checkmark (s) & - & - & \checkmark & \checkmark (c) & - & -\\
        \cite{Wu2018} & \checkmark & - & - & - & \checkmark & \checkmark (c) & - & \checkmark (H)\\
        \cite{Zengen2020} & \checkmark & - & - & - & - & \checkmark (d) & - & -\\

        \cite{Bai2021} & \checkmark & \checkmark (m) & - & - & - & - & - & \checkmark (H)\\
        \cite{Alsurdeh2021} & \checkmark & - & - & - & - & - & \checkmark & \checkmark (H)\\
        \cite{Stavrinides2019} & \checkmark & - & - & - & - & - & \checkmark & \checkmark (H)\\
    
        This work                  & \checkmark & \checkmark (m) & \checkmark & \checkmark & \checkmark & \checkmark (c) & \checkmark & \checkmark (H)\\
        \bottomrule
        \multicolumn{9}{l}{${}^{1}$\underline{s}ingle/\underline{m}ultiple capabilities (e.g., sensors and/or actuators) per CPS device.}\\
        \multicolumn{9}{l}{${}^{2}$\underline{d}iscrete-time/\underline{c}ontinuous-time approach. \quad ${}^{3}$\underline{h}omogeneous/\underline{H}eterogeneous processors.}\\
    \end{tabular}
}
\vspace{-9pt}
\end{table}

In addition to exact approaches, heuristic techniques have also received significant attention for workflow scheduling in CPS and multi-tier architectures \cite{Bai2021, Alsurdeh2021, Stavrinides2019}. For instance, the heuristic in \cite{Bai2021} aims to schedule tasks with different sensing/actuating requirements in a CPS with multiple sensors and actuators per device.
Moreover, a two-stage strategy is proposed in \cite{Alsurdeh2021}, combining a heuristic search approach and task clustering to schedule workflow applications in an edge-cloud environment. In the same context, a two-phase heuristic is presented in \cite{Stavrinides2019}, which makes scheduling decisions based on the computation and communication requirements of each task in a three-tier system.
Even though these methods are suitable for heterogeneous multicore processors and workflows with execution deadlines, they do not consider the specific environment, nor all the constraints examined in this work.

Among the most well-established and effective workflow scheduling heuristics is the heterogeneous earliest finish time (HEFT) \cite{Topcuoglu2002}. It remains one of the most frequently used techniques for distributed heterogeneous (multicore) processors, providing high-quality schedules \cite{Kuhbacher2019}. For example, it is employed in \cite{Aldegheri2020} to schedule embedded vision applications in a multicore architecture. Similarly, it is used in \cite{Kuhbacher2019} to schedule dataflow applications in a safety-critical embedded system. 
However, HEFT does not support the deadline, capability, memory, storage, and energy constraints arising from the system setting considered in this work.
Moreover, as it is a heuristic method, it cannot guarantee optimal solutions.

Overall, existing exact and heuristic workflow scheduling methods do not comprehensively address all the constraints considered in the proposed MILP approach, nor examine the specific CPS architecture. The qualitative comparison of this work with the presented research efforts is summarized in \Cref{table:comparison}.
Although we previously investigated edge-hub-cloud systems in \cite{Kouloumpris2019, Kouloumpris2020, Kouloumpris2019b, Kouloumpris2024}, we did not consider multiple edge devices, nor a CPS with different sensing, actuating, or other specialized capabilities. More importantly, our previous approaches pose a key limitation, as they can only determine where to allocate tasks, but not when to execute them, performing only the mapping step of the scheduling process.

\section{Proposed MILP Approach}
\label{sec:framework}

\subsection{Application Model}
\label{subsec:appModel}
The workflow application comprises a set of $\alpha$ tasks $\mathcal{T} = \{ \tau_i \,|\, 1 \leq i \leq \alpha \}$. It is represented by a task graph (TG) in the form of a directed acyclic graph $G=( \mathcal{N, \mathcal{A}})$ \cite{Aldegheri2020}. $\mathcal{N}=\{ N_i \,|\, \tau_i \in \mathcal{T} \}$ is the set of its nodes, whereas $\mathcal{A} = \{ A_{i \rightarrow j} \, | \, N_{i}, \allowbreak N_{j} \in \mathcal{N}, \, i \neq j, \, \exists \text{ a data dependency } N_{i} \rightarrow N_{j} \}$ is the set of its arcs.
A node $N_{i} \in \mathcal{N}$ represents a task $\tau_i \in \mathcal{T}$. 
A task is an indivisible unit of work.
Tasks are considered to be non-preemptive, as preemption in time-constrained applications may lead to performance degradation \cite{Baek2020}.
An arc $A_{i \rightarrow j} \in \mathcal{A}$ between two nodes $N_{i}$ and $N_{j}$ (corresponding to parent task $\tau_i$ and child task $\tau_j$, respectively), denotes the communication and precedence relationship between the two tasks.
There is a predefined deadline (i.e., latency threshold) $L_{\mathrm{thr}}$ before which all tasks of the application should be completed. $L_{\mathrm{thr}}$ also denotes the time horizon of the examined scheduling problem.

\subsection{System Model}
\label{subsec:sysModel}
We consider a CPS comprising a set of $\beta$ different devices $\mathcal{U} = \{ u_{\lambda k} \,| \allowbreak \, \lambda \in \{\mathrm{e, h, c}\}, \allowbreak \, 1 \leq k \leq \beta_{\lambda}, \allowbreak \, \sum_{\lambda \in \{ \mathrm{e}, \mathrm{h}, \mathrm{c} \}} \beta_{\lambda} = \beta \}$. $\lambda$ denotes the type of the device (e represents an edge device, h a hub device, and c a cloud server), whereas $k$ and $\beta_{\lambda}$ are the device index and the number of devices of the particular type, respectively. 
All devices in $\mathcal{U}$ are multicore. For the execution of the specific application, we reserve on each device $u_{\lambda k} \in \mathcal{U}$ a set of $\gamma_{\lambda k}$ processing cores $\mathcal{P}_{\lambda k} = \{ p_{\lambda k.q}\,| \allowbreak \, 1 \leq q \leq \gamma_{\lambda k} \}$ ($q$ denotes the processing core index). 
The set of all reserved processing cores on all system devices is defined as $\mathcal{P} = \bigcup_{u_{\lambda k} \in \mathcal{U}} \mathcal{P}_{\lambda k}$.
The processing cores in $\mathcal{P}$ may be heterogeneous. 
Thus, the latency and power consumption required to execute a given task may vary.
Each processing core can execute one task at a time. 
As the system resources are shared with other applications, for the execution of the examined workflow we consider memory, storage, and energy budgets for each device  $u_{\lambda k}$. These budgets, denoted by $M_{\lambda k}^{\mathrm{bgt}}$, $S_{\lambda k}^{\mathrm{bgt}}$, and $E_{\lambda k}^{\mathrm{bgt}}$, respectively, are shared among the reserved processing cores of each device.

Furthermore, we consider a set of $\delta$ device capabilities $\mathcal{C} = \{ c_a \,| \allowbreak \, 0 \leq a < \delta \}$. $c_0$ represents the basic computational capability of a device, whereas $c_a$ with $a > 0$ denotes a specialized capability (in addition to the device's basic computational capability), such as a specific sensor, actuator, or software/hardware module.
Each device $u_{\lambda k} \in \mathcal{U}$ has a set of capabilities $\mathcal{C}_{\lambda k} \subseteq \mathcal{C}$, such that  $c_0 \in \mathcal{C}_{\lambda k}$. 
The binary parameter $y_{\lambda k}^{c_a}$ indicates whether device $u_{\lambda k}$ features capability $c_a \in  \mathcal{C}$ ($y_{\lambda k}^{c_a} = 1$) or not ($y_{\lambda k}^{c_a}=0$).
Accordingly, the binary parameter $z_i^{c_a}$ denotes if task $\tau_i$ requires capability $c_a \in  \mathcal{C}$  ($z_i^{c_a} = 1$) or not ($z_i^{c_a} = 0$). 
We consider that each task $\tau_i \in \mathcal{T}$ requires one device capability \cite{Bai2021}.

Each pair of devices can communicate directly or indirectly through another device. Specifically, the communication channel between devices $u_{\lambda k}$ and $u_{\mu l}$ is defined as $\xi_{\lambda k, \mu l} = \langle \theta_{\lambda k, \mu l}, \allowbreak \pi_{\lambda k, \mu l}, \allowbreak \rho_{\lambda k, \mu l}, \allowbreak \mathcal{I}_{\lambda k, \mu l} \rangle$, where $\theta_{\lambda k, \mu l}$ denotes its bandwidth, whereas $\pi_{\lambda k, \mu l}$ and $\rho_{\lambda k, \mu l}$ denote the energy consumption for the transmission and reception of a unit of data over the particular channel, respectively. 
If $u_{\lambda k}$ and $u_{\mu l}$ can only communicate through device $u_{\nu m}$, then $\mathcal{I}_{\lambda k, \mu l}=\{ u_{\nu m} \}$, otherwise $\mathcal{I}_{\lambda k, \mu l}= \varnothing$.

\subsection{Extended Task Graph}
\label{subsec:etg}
To facilitate the MILP formulation of the considered problem, we transform the initial TG $G=(\mathcal{N, \mathcal{A}})$ of the application into an extended task graph (ETG) $G^{\prime}=(\mathcal{N}^{\prime}, \mathcal{A}^{\prime})$, based on the main principles of our approach in \cite{Kouloumpris2019}. The ETG encapsulates both the application and system models, incorporating their communication and energy aspects.
In particular, each node $N_i \in \mathcal{N}$ in $G$ is transformed into a composite node $N_i^{\prime} \in \mathcal{N}^{\prime}$ in $G^{\prime}$, such that:
\begin{equation}
\label{eq:nodeTransformation}
 N_i^{\prime} = \{ N_{i \lambda k.q} \, | \, p_{\lambda k.q} \in \mathcal{P}, \, y_{\lambda k}^{c_a}\,z_i^{c_a} = 1, \, c_a \in \mathcal{C} \}.
\end{equation}
Hence, $N_i^{\prime}$ is the set of nodes corresponding to the possible allocations of task $\tau_i$ on the reserved processing cores of the
devices featuring its required capability.
An individual node $N_{i \lambda k.q } \in N_i^{\prime}$ (denoting a specific allocation of $\tau_i$) is referred to as a \emph{candidate node}.

Accordingly, each arc $A_{i \rightarrow j} \in \mathcal{A}$ in $G$ is transformed into a composite arc $A^{\prime}_{i \rightarrow j} \in \mathcal{A}^{\prime}$ in $G^{\prime}$, such that:
\begin{equation}
 \label{eq:arcTransformation}
 \begin{split}
     A^{\prime}_{i \rightarrow j}  \hspace{-3pt} = &  \{ A_{i \lambda k.q \rightarrow j \mu l.r} \hspace{-2pt} = \hspace{-2pt} N_{i \lambda k.q} \hspace{-2pt} \rightarrow \hspace{-2pt} N_{j \mu l.r} \, | \, p_{\lambda k.q}, p_{\mu l.r} \hspace{-1.5pt} \in \hspace{-1.5pt} \mathcal{P}\hspace{-1.25pt}, \\
     & \quad  \, y_{\lambda k}^{c_a}\,z_i^{c_a} \hspace{-2pt} = \hspace{-2pt} y_{\mu l}^{c_b}\,z_j^{c_b} \hspace{-2pt} = \hspace{-2pt} 1, \, c_a, c_b \in \mathcal{C}\},
\end{split}
\end{equation}
where an individual arc $A_{i \lambda k.q \rightarrow j \mu l.r} \in A^{\prime}_{i \rightarrow j}$ represents the data flow between candidate nodes $N_{i \lambda k.q}$ and $N_{j \mu l.r}$.

\begin{figure}[!t]
    \centering
    \includegraphics[width=0.75\columnwidth]{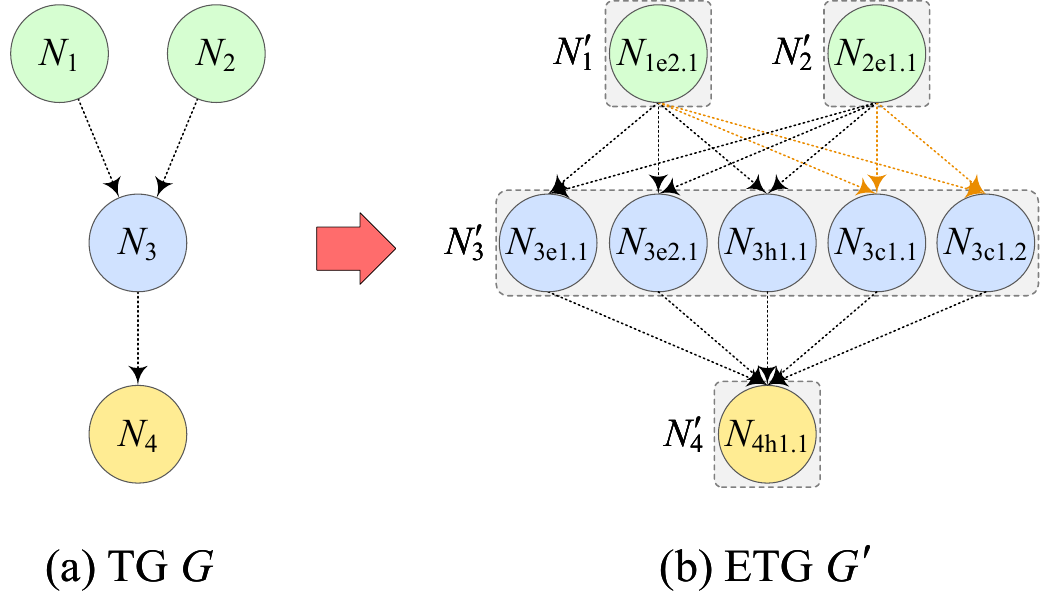}
    \caption{Transformation example: (a) the initial TG and (b) the final ETG.}
    \label{fig:example}
    \vspace{-5pt}
\end{figure}

An example of transforming a TG $G$ into ETG $G^{\prime}$ is illustrated in \cref{fig:example}. $G$ consists of four tasks/nodes, with tasks $N_1$, $N_2$, $N_3$, and $N_4$ requiring device capabilities $c_2$, $c_1$, $c_0$, and $c_3$, respectively (entry/exit tasks are depicted in green/yellow and intermediate tasks in blue). The underlying edge-hub-cloud CPS comprises two edge devices $u_{\mathrm{e}1}$ and $u_{\mathrm{e}2}$ with capabilities $\mathcal{C}_{\mathrm{e}1} = \{ c_0, c_1\}$ and $\mathcal{C}_{\mathrm{e}2} = \{ c_0, c_2\}$, respectively, a hub device $u_{\mathrm{h}1}$ with capabilities $\mathcal{C}_{\mathrm{h}1} = \{ c_0, c_3\}$, and a cloud server $u_{\mathrm{c}1}$ with capability $\mathcal{C}_{\mathrm{c}1} = \{ c_0\}$. For the execution of $G$, one processing core is reserved on each of the devices $u_{\mathrm{e}1}$, $u_{\mathrm{e}2}$, and $u_{\mathrm{h}1}$, and two processing cores on $u_{\mathrm{c}1}$.        
Devices $u_{\mathrm{e}1}$ and $u_{\mathrm{e}2}$ utilize $u_{\mathrm{h}1}$ to communicate with $u_{\mathrm{c}1}$. Based on the capability requirements of each task, and the capabilities featured by each device, the possible allocations of the tasks on the processing cores result in the generation of eight candidate nodes in $G^{\prime}$ (one candidate node for $N_1$, $N_2$, and $N_4$, and five for $N_3$). 
Each arc between two tasks in $G$ is transformed into a set of arcs in $G^{\prime}$ connecting all pairs of candidate nodes of the two tasks. 
The arcs in $G^{\prime}$ that involve indirect communication between devices are shown in orange.

In the worst case of our transformation technique, where all devices feature all capabilities in $\mathcal{C}$, for $\gamma$ processing cores the number of candidate nodes in ETG $G^{\prime}$ increases by $\gamma$ times, and the number of arcs by $\gamma^2$ times, compared to those in TG $G$. 
For applications like the ones motivating this work \cite{Zheng2021, Alam2017, Kashino2019, Savva2021}, which do not typically have an excessive number of tasks, the optimality of our approach outweighs the moderate complexity introduced by the increase in task graph size. This is demonstrated experimentally in \cref{subsec:scalability}.

\subsubsection{Candidate Node Parameters}
\label{subsubsec:nodeParams}
In addition to the binary parameters $y_{\lambda k}^{c_a}$ and $z_i^{c_a}$ defined in Section \ref{subsec:sysModel}, a candidate node $N_{i \lambda k.q} \in G^{\prime}$ has the following parameters:
\vspace{-5pt}
    \begin{itemize}
        \item $D_i$: output data size of task $\tau_i$.
        
        \item $M_i$: main memory required by $\tau_i$.
        
        \item $S_i$: storage required by $\tau_i$.

        \item $\mathcal{Q}_i$: set of child tasks of $\tau_i$.
        \item $L_{i \lambda k.q}$: execution time of $\tau_i$ on processing core $p_{\lambda k.q}$.
       
        \item $P_{i \lambda k.q}$: power required to execute $\tau_i$ on $p_{\lambda k.q}$. 
        
        \item $E_{i \lambda k.q}$: energy required to execute $\tau_i$ on $p_{\lambda k.q}$, given by:
        \vspace{-15pt}
        \begin{equation}
        \label{eq:compEnergy}
            E_{i \lambda k.q} = L_{i \lambda k.q} \, P_{i \lambda k.q}.
        \end{equation}
    \end{itemize}

\vspace{-5pt}
\subsubsection{Arc Parameters}
\label{subsubsec:arcParams}
An arc $A_{i \lambda k.q \rightarrow j \mu l.r} \in G^{\prime}$ has the following parameters:
\vspace{-5pt}
\begin{itemize}
    \item  $\sigma_{i \lambda k.q \rightarrow j \mu l.r}^{\nu m}$: binary parameter denoting whether  $A_{i \lambda k.q \rightarrow j \mu l.r}$ involves indirect communication between devices $u_{\lambda k}$ and $u_{\mu l}$ through device $u_{\nu m}$:
    \vspace{-10pt}
    \begin{equation}
    \label{eq:virtualNodeParam}
        \sigma_{i \lambda k.q \rightarrow j \mu l.r}^{\nu m} = 
        \begin{cases}
            1, & \text{if $\mathcal{I}_{\lambda k, \mu l}=\{ u_{\nu m} \}$},\\
            0, & \text{if $\mathcal{I}_{\lambda k, \mu l}= \varnothing$}.
        \end{cases}
    \end{equation}

    \item $CL_{i \lambda k.q \rightarrow j \mu l.r}$: time required to transfer the output data $D_{i}$ of task $\tau_i$ that is allocated on $p_{\lambda k.q}$, to task $\tau_j$ that is allocated on $p_{\mu l.r}$:
    \vspace{-10pt}
    \begin{equation}
    \label{eq:commLatency}
        \hspace{-19pt} CL_{i \lambda k.q \rightarrow j \mu l.r} =  
        \begin{cases}
             \hspace{-2pt} \frac {D_i}{\theta_{\lambda k, \mu l}}, & \hspace{-65pt} \text{if $\sigma_{i \lambda k.q \rightarrow j \mu l.r}^{\nu m} = 0$},\\
             & \hspace{-65pt} \text{$(\lambda, k) \neq (\mu, l)$},\\
             \hspace{-2pt} D_i \left( \frac{1}{\theta_{\lambda k, \nu m}} + \frac{1}{\theta_{\nu m, \mu l}} \right), &\\
             & \hspace{-65pt} \text{if $\sigma_{i \lambda k.q \rightarrow j \mu l.r}^{\nu m} = 1$},\\
             \hspace{-2pt} 0, & \hspace{-65pt} \text{if $(\lambda, k) = (\mu, l)$}.
        \end{cases}
    \end{equation}

    \item $CE_{i \lambda k.q \rightarrow j \mu l.r}$: energy required to transfer $D_{i}$ from task $\tau_i$ (allocated on $p_{\lambda k.q}$) to task $\tau_j$ (allocated on $p_{\mu l.r}$):
    \vspace{-5pt}
    \begin{equation}
    \label{eq:commEnergy}
        \hspace{-19pt} CE_{i \lambda k.q \rightarrow j \mu l.r} = 
        \begin{cases}
             \hspace{-2pt} D_i \left( \pi_{\lambda k, \mu l} + \rho_{\lambda k, \mu l} \right), &\\
             \hspace{27pt} \text{if $\sigma_{i \lambda k.q \rightarrow j \mu l.r}^{\nu m} = 0$},&\\ \hspace{27pt} \text{$(\lambda, k) \neq (\mu, l)$},\\
             \hspace{-2pt} D_i ( \pi_{\lambda k, \nu m} + \rho_{\lambda k, \nu m}\\ 
             \hspace{15pt} + \, \pi_{\nu m, \mu l} + \rho_{\nu m, \mu l} ), &\\
             \hspace{27pt} \text{if $\sigma_{i \lambda k.q \rightarrow j \mu l.r}^{\nu m} = 1$},\\
             \hspace{-2pt} 0, & \hspace{-91pt} \text{if $(\lambda, k) = (\mu, l)$}.
        \end{cases}
    \end{equation} 
\end{itemize}

\subsection{Problem Formulation}
\label{subsec:milp}
We leverage the resulting ETG $G^{\prime}$ to formulate the scheduling problem as a continuous-time MILP model, as follows.

\subsubsection{Decision Variables}
\label{subsubsec:variables}
We employ the following variables:
\begin{itemize}
    \item $x_{i \lambda k.q}$: binary variable corresponding to a candidate node $N_{i \lambda k.q} \in G^{\prime}$, such that $x_{i \lambda k.q} = 1$ if $N_{i \lambda k.q}$ is selected, and $x_{i \lambda k.q} = 0$ otherwise.
   
    \item $x_{i \lambda k.q \rightarrow j \mu l.r}$: binary variable corresponding to an arc $A_{i \lambda k.q \rightarrow j \mu l.r} \in G^{\prime}$, such that $x_{i \lambda k.q \rightarrow j \mu l.r} = 1$ if $A_{i \lambda k.q \rightarrow j \mu l.r}$ is selected, and $x_{i \lambda k.q \rightarrow j \mu l.r} = 0$ otherwise.

    \item $t_i$: continuous variable denoting the start time of task $\tau_i$, corresponding to the set $N_i^{\prime} \in G^{\prime}$. 

    \item $T$: continuous variable denoting the completion time of the application.

    \item $x_{i,j}$: auxiliary binary variable denoting whether task $\tau_i$ will be executed before task $\tau_j$ ($x_{i,j}=1$) or not ($x_{i,j}=0$). It is used when both tasks are allocated on the same processing core and do not have a precedence relationship between them, and thus their order of execution should be determined to avoid any overlap.

    \item $x_n^{i \lambda k.q}$: auxiliary binary variable denoting whether task $\tau_i$ will be in execution on its allocated processing core $p_{\lambda k.q}$ at time $e_n$ ($x_n^{i \lambda k.q} = 1$) or not ($x_n^{i \lambda k.q} = 0$). It is used to prevent exceeding the capability, memory, and storage capacity of a device during the execution of its assigned tasks (i.e., it is used to formulate time-dependent cumulative constraints). For this purpose, taking into account that the time interval in which a task $\tau_i$ will be in execution on $p_{\lambda k.q}$ is given by $[t_i, t_i + L_{i \lambda k.q})$, we consider a set comprising the start times of all tasks, i.e., $\mathcal{E} = \{e_n \,| \, e_n = t_i, \allowbreak \, N_i^{\prime} \in \mathcal{N}^{\prime} \}$. 
    To preserve the linearity of the model, an additional related auxiliary binary variable $\hat{x}_n^{i \lambda k.q}$ is used.
\end{itemize}

\subsubsection{Objective \& Constraints}
\label{subsubsec:objective}
Our aim is to minimize the overall latency (i.e., the completion time) of the application:
\begin{equation}
\label{eq:objective}
    \min T
\end{equation}
subject to the following constraints.

\paragraph{Candidate node selection constraints} Only one candidate node per task should be selected in $G^{\prime}$, i.e., each task should be allocated on only one processing core: 
\begin{equation}
\label{eq:one}
    \sum_{N_{i \lambda k.q} \in N_i^{\prime}} \hspace{-10pt} x_{i \lambda k.q} =  1, \, \forall \, N_i^{\prime} \in \mathcal{N}^{\prime}.
\end{equation}

\paragraph{Arc selection constraints}
If a parent and child candidate nodes are selected, their corresponding arc should be selected as well:
\begin{equation}
\label{eq:three1}
    x_{i \lambda k.q \rightarrow j \mu l.r} \leq x_{i \lambda k.q}, \, \forall \, A_{i \lambda k.q \rightarrow j \mu l.r} \in \mathcal{A}^{\prime},
\end{equation}
\begin{equation}
\label{eq:three2}
    x_{i \lambda k.q \rightarrow j \mu l.r} \leq x_{j \mu l.r}, \, \forall \, A_{i \lambda k.q \rightarrow j \mu l.r} \in \mathcal{A}^{\prime},
\end{equation}
\begin{equation}
\label{eq:three3}
    x_{i \lambda k.q \rightarrow j \mu l.r} \geq x_{i \lambda k.q} + x_{j \mu l.r} - 1, \, \forall \, A_{i \lambda k.q \rightarrow j \mu l.r} \in \mathcal{A}^{\prime}.
\end{equation}

\paragraph{Task precedence constraints} The precedence relationships among the tasks should be preserved:
\begin{equation}
\label{eq:four}
    \begin{split}
        t_i + L_{i \lambda k.q}\,x_{i \lambda k.q} + CL_{i \lambda k.q \rightarrow j \mu l.r}\,x_{i \lambda k.q \rightarrow j \mu l.r} \leq t_j, \\
        \forall A_{i \lambda k.q \rightarrow j \mu l.r} \in \mathcal{A}^{\prime}. 
    \end{split}
\end{equation}

\paragraph{Application completion time \& deadline constraints} The completion time of the application should be equal to the completion time of its last task\,\eqref{eq:five}, and within the predefined deadline\,\eqref{eq:six}:
\begin{equation}
\label{eq:five}
    t_i + L_{i \lambda k.q}\,x_{i \lambda k.q} \leq T, \, \forall \, N_{i \lambda k.q} \in \mathcal{N}^{\prime},     
\end{equation}
\begin{equation}
\label{eq:six}
    T \leq L_{\mathrm{thr}}.
\end{equation}

\paragraph{Task non-overlapping constraints} Two tasks without a precedence relationship between them, allocated on the same processing core, should not be executed at the same time, as each processing core can process only one task at a time:
\begin{equation}
\label{eq:eight1}
\begin{split}
    t_i + L_{i \lambda k.q}\,x_{i \lambda k.q} \leq t_j + \left( 3 - x_{i \lambda k.q} - x_{j \lambda k.q} - x_{i,j} \right) \mathit{\Omega},\\ 
    \forall \, N_{i \lambda k.q}, N_{j \lambda k.q} \in \mathcal{N}^{\prime}, \, i < j, \, \tau_i \notin \mathcal{Q}_j, \, \tau_j \notin \mathcal{Q}_i,  
\end{split}    
\end{equation}
\begin{equation}
\label{eq:eight2}
\begin{split}
    t_j + L_{j \lambda k.q}\,x_{j \lambda k.q} \leq t_i + \left( 2 - x_{i \lambda k.q} - x_{j \lambda k.q} + x_{i,j} \right) \mathit{\Omega},\\
    \forall \, N_{i \lambda k.q}, N_{j \lambda k.q} \in \mathcal{N}^{\prime}, \, i < j, \, \tau_i \notin \mathcal{Q}_j, \, \tau_j \notin \mathcal{Q}_i.  
\end{split}    
\end{equation}

\paragraph{Task execution constraints} The time instants $e_n \in \mathcal{E}$ at which each task will be in execution on its allocated processing core are determined by the following constraints:
\begin{equation}
\label{eq:nine}
    x_n^{i \lambda k.q} \leq x_{i \lambda k.q}, \, \forall \, e_n \in \mathcal{E}, \, \forall \, N_{i \lambda k.q} \in \mathcal{N}^{\prime},
\end{equation}
\vspace{1pt}
\begin{equation}
\label{eq:ten1}
    t_i \leq e_n + \left( 2 - x_{i \lambda k.q} - x_n^{i \lambda k.q} \right) \mathit{\Omega}, \, \forall \, e_n \in \mathcal{E}, \, \forall \, N_{i \lambda k.q} \in \mathcal{N}^{\prime},
\end{equation}
\vspace{1pt}
\begin{equation}
\label{eq:ten2}
\begin{split}
    e_n + \epsilon \leq t_i + L_{i \lambda k.q}\,x_{i \lambda k.q} + \left( 2 - x_{i \lambda k.q} - x_n^{i \lambda k.q} \right) \mathit{\Omega},\\ 
    \forall \, e_n \in \mathcal{E}, \, \forall \, N_{i \lambda k.q} \in \mathcal{N}^{\prime},
\end{split}
\end{equation}
\vspace{1pt}
\begin{equation}
\label{eq:ten3}
\begin{split}
    e_n + \epsilon \leq t_i + \left( 2 - x_{i \lambda k.q} + x_n^{i \lambda k.q} - \hat{x}_n^{i \lambda k.q} \right) \mathit{\Omega},\\ 
    \forall \, e_n \in \mathcal{E}, \, \forall \, N_{i \lambda k.q} \in \mathcal{N}^{\prime},
\end{split}
\end{equation}
\vspace{1pt}
\begin{equation}
\label{eq:ten4}
\begin{split}
    t_i + L_{i \lambda k.q}\,x_{i \lambda k.q} \leq e_n + \left( 1 - x_{i \lambda k.q} + x_n^{i \lambda k.q} + \hat{x}_n^{i \lambda k.q} \right) \mathit{\Omega},\\
    \forall \, e_n \in \mathcal{E}, \, \forall \, N_{i \lambda k.q} \in \mathcal{N}^{\prime}.
\end{split}
\end{equation}

\paragraph{Device capability, memory \& storage constraints} At each time instant $e_n \in \mathcal{E}$, no more than one task among those being executed on the processing cores of a particular device, should require a specific specialized capability\,\eqref{eq:eleven}.
Similarly, at each time instant $e_n \in \mathcal{E}$ the memory\,\eqref{eq:twelve} and storage\,\eqref{eq:thirteen} budgets of each device should not be exceeded:
\begin{equation}
\label{eq:eleven}
\begin{split}
    \sum_{N_{i \lambda k.q} \in \mathcal{N}^{\prime}} \hspace{-10pt} x_n^{i \lambda k.q}\,y_{\lambda k}^{c_a}\,z_i^{c_a} \hspace{-1pt} \leq \hspace{-1.5pt} 1,
    \hspace{0.75pt} \forall \hspace{0.75pt} e_n \hspace{-1.25pt} \in \hspace{-1pt} \mathcal{E},\hspace{0.75pt} \forall \hspace{0.75pt} u_{\lambda k} \hspace{-1.25pt} \in \hspace{-1pt} \mathcal{U}, \hspace{0.75pt} \forall \hspace{0.75pt} c_a \hspace{-1.25pt} \in \hspace{-1pt} \mathcal{C}, a \hspace{-1.25pt} > \hspace{-1.25pt} 0,
\end{split}
\end{equation}
\begin{equation}
\label{eq:twelve}
    \sum_{N_{i \lambda k.q} \in \mathcal{N}^{\prime}} \hspace{-10pt} M_i\,x_n^{i \lambda k.q} \leq M_{\lambda k}^{\mathrm{bgt}}, \, \forall \, e_n \in \mathcal{E}, \, \forall \, u_{\lambda k} \in \mathcal{U},
\end{equation}
\begin{equation}
\label{eq:thirteen}
    \sum_{N_{i \lambda k.q} \in \mathcal{N}^{\prime}} \hspace{-10pt} S_i\,x_n^{i \lambda k.q} \leq S_{\lambda k}^{\mathrm{bgt}}, \, \forall \, e_n \in \mathcal{E}, \, \forall \, u_{\lambda k} \in \mathcal{U}.
\end{equation}

\paragraph{Device energy constraints} The energy budget of each device should not be exceeded for the execution of the application, considering both the computational and communication energy requirements of the tasks.
Regarding communication energy, we consider all data transmitted from and received at each device (either directly or indirectly to/from another device), including the case where a device is used for the communication between other devices:
\begin{equation}
\label{eq:fourteen}
\begin{split}
    & \sum_{N_{i \lambda k.q} \in \mathcal{N}^{\prime}} \hspace{-10pt} E_{i \lambda k.q}\,x_{i \lambda k.q}\\
    & + \hspace{-10pt} \sum_{A_{i \lambda k.q \rightarrow j \mu l.r} \in \mathcal{A}^{\prime}} \hspace{-20pt} D_i \, x_{i \lambda k.q \rightarrow j \mu l.r} \big( \big. \pi_{\lambda k, \mu l} \left( 1 - \sigma_{i \lambda k.q \rightarrow j \mu l.r}^{\nu m} \right)\\
    & \hspace{118pt} + \pi_{\lambda k, \nu m} \, \sigma_{i \lambda k.q \rightarrow j \mu l.r}^{\nu m} \big) \big. \\ 
    & + \hspace{-10pt} \sum_{A_{j \mu l.r \rightarrow i \lambda k.q} \in \mathcal{A}^{\prime}} \hspace{-20pt} D_j \, x_{j \mu l.r \rightarrow i \lambda k.q} \big( \big. \rho_{\mu l, \lambda k} \left( 1 - \sigma_{j \mu l.r \rightarrow i \lambda k.q}^{\nu m} \right)\\
    & \hspace{118pt} + \rho_{\nu m, \lambda k} \, \sigma_{j \mu l.r \rightarrow i \lambda k.q}^{\nu m} \big) \big. \\
    & + \hspace{-10pt} \sum_{A_{i \mu l.q \rightarrow j \nu m.r} \in \mathcal{A}^{\prime}} \hspace{-20pt} D_i \, x_{i \mu l.q \rightarrow j \nu m.r} \left( \rho_{\mu l, \lambda k} + \pi_{\lambda k, \nu m} \right) \sigma_{i \mu l.q \rightarrow j \nu m.r}^{\lambda k}\\
    & \leq E_{\lambda k}^{\mathrm{bgt}}, \, \forall \, u_{\lambda k} \in \mathcal{U}, \, (\lambda, k) \neq (\mu, l) \neq (\nu, m).
\end{split}
\end{equation}

\paragraph{Binary \& non-negativity constraints} 
The binary nature\,\eqref{eq:fifteen}--\eqref{eq:eighteen} and non-negativity\,\eqref{eq:nineteen}, \eqref{eq:twenty} of the binary and continuous decision variables, respectively, should be ensured:
\vspace{-8pt}
\begin{equation}
\label{eq:fifteen}
     x_{i \lambda k.q} \in \{ 0, 1 \}, \, \forall \, N_{i \lambda k.q} \in \mathcal{N}^{\prime},
\end{equation}
\begin{equation}
\label{eq:sixteen}
     x_{i \lambda k.q \rightarrow j \mu l.r} \in \{ 0, 1 \}, \, \forall \, A_{i \lambda k.q \rightarrow j \mu l.r} \in \mathcal{A}^{\prime}, 
\end{equation}
\begin{equation}
\label{eq:seventeen}
    x_{i,j} \hspace{-0.5pt} \in \hspace{-0.5pt} \{ 0, 1 \}\hspace{-0.25pt}, \hspace{0.75pt} \forall \hspace{0.75pt} N_{i \lambda k.q}, N_{j \lambda k.q} \hspace{-0.5pt} \in \hspace{-0.5pt} \mathcal{N}^{\prime}, \hspace{0.75pt} i \hspace{-0.5pt} < \hspace{-0.5pt} j\hspace{-0.25pt}, \hspace{0.75pt} \tau_i \hspace{-0.5pt} \notin \hspace{-0.5pt} \mathcal{Q}_j, \hspace{0.75pt} \tau_j \hspace{-0.5pt} \notin \hspace{-0.5pt} \mathcal{Q}_i,
\end{equation}
\begin{equation}
\label{eq:eighteen}
    x_n^{i \lambda k.q}, \hat{x}_n^{i \lambda k.q} \in \{ 0, 1 \}, \, \forall \, e_n \in \mathcal{E}, \, \forall \, N_{i \lambda k.q} \in \mathcal{N}^{\prime},
\end{equation}
\begin{equation}
\label{eq:nineteen}
    t_i \geq 0, \, \forall \, N_i^{\prime} \in \mathcal{N}^{\prime},
\end{equation}
\begin{equation}
\label{eq:twenty}
     T \geq 0.
\end{equation}

\vspace{-5pt}
It is noted that in \eqref{eq:ten2} and \eqref{eq:ten3}, we utilize a positive tolerance constant $\epsilon$, sufficiently smaller than the variables and parameters used in the model, to convert the constraints to non-strict inequalities, as strict inequalities are not supported in MILP. Furthermore, in \eqref{eq:eight1}, \eqref{eq:eight2}, \eqref{eq:ten1}--\eqref{eq:ten4}, we employ a constant $\mathit{\Omega}$, sufficiently larger than the worst-case completion time of any task of the application, to formulate the conditional aspects of the constraints in linear form. For example, \eqref{eq:eight1} becomes meaningful (i.e., $ t_i + L_{i \lambda k.q} \leq t_j$) only if tasks $\tau_i$ and $\tau_j$ are allocated on the same processing core $p_{\lambda k.q}$ (i.e., $x_{i \lambda k.q} = x_{j \lambda k.q} = 1$) and $\tau_i$ is executed before $\tau_j$ (i.e., $x_{i,j} = 1$). Otherwise, $\mathit{\Omega}$ remains on the right-hand side of \eqref{eq:eight1}, forcing it to be always true. Constraints \eqref{eq:eight2}, \eqref{eq:ten1}--\eqref{eq:ten4} are modeled in a similar fashion.

\section{HEFT Extension}
\label{sec:heft}
As showcased in \cref{sec:related}, our MILP approach is the first to optimally solve the examined problem in the considered CPS under the specific constraints. Thus, any modifications to other exact methods to account for the particular constraints would inevitably lead to a formulation identical to ours. Therefore, we chose to compare our approach with HEFT \cite{Topcuoglu2002, Kuhbacher2019, Aldegheri2020}, which is one of the most frequently used and effective scheduling heuristics for workflow applications.
HEFT involves two phases, (a) a task prioritization phase, where tasks are prioritized according to their upward rank (i.e., the longest distance to an exit task, in terms of latency), and (b) a processing core selection phase, where each task, in order of priority, is allocated to the processing core that can provide it with the minimum finish time, utilizing any schedule gaps.

As HEFT  does not inherently support all the constraints of the proposed MILP approach, we extended it to ensure a meaningful and fair comparison.
Specifically, we adapted HEFT to leverage the ETG $G^{\prime}$ of the application (rather than TG $G$), and enhanced it by incorporating into its second phase the deadline \eqref{eq:six}, capability \eqref{eq:eleven}, memory \eqref{eq:twelve}, storage \eqref{eq:thirteen}, and energy \eqref{eq:fourteen} constraints considered in our method (the original version of HEFT \cite{Topcuoglu2002} supports only the precedence \eqref{eq:four} and non-overlapping \eqref{eq:eight1}, \eqref{eq:eight2} constraints).
As the enhanced version of HEFT (shown in \cref{alg:heft}) takes as input the ETG $G^{\prime}$, its first phase is converted into a candidate node prioritization phase (lines 1--5), and its second phase into a candidate node selection phase (lines 6--59).

\newlength{\oldtextfloatsep}
\setlength{\oldtextfloatsep}{\textfloatsep}

\setlength{\textfloatsep}{10pt}

\begin{algorithm}[!t]

\scriptsize

\SetInd{0.65em}{0.65em} 

\KwIn{ETG $G^{\prime}=(\mathcal{N}^{\prime}, \mathcal{A}^{\prime})$.}
\KwOut{$x_{i \lambda k.q} \, \forall \, N_{i \lambda k.q} \in \mathcal{N}^{\prime}$,
$x_{i \lambda k.q \rightarrow j \mu l.r} \, \forall \, A_{i \lambda k.q \rightarrow j \mu l.r} \in \mathcal{A}^{\prime}$,
and $t_i \, \forall \, N_i^{\prime} \in G^{\prime}$.}

\tcp{Phase 1 - candidate node prioritization:}

\ForEach{\textup{candidate node \hspace{-4pt} $N_{i \lambda k.q} \hspace{-2pt} \in \hspace{-2pt} \mathcal{N}^{\prime}$ \hspace{-4pt} starting from exit task candidate nodes}}{
Calculate upward rank $R_{i \lambda k.q}$ using \eqref{eq:rank}\;
}

$\mathit{\Lambda} \gets \mathcal{N}^{\prime}$\;
Sort list $\mathit{\Lambda}$ by non-increasing order of $R_{i \lambda k.q}$\;

\tcp{Phase 2 - candidate node selection:}

$isInfeasible \gets 0$\;
$\mathcal{N}^{\prime}_{\mathrm{sel}} \gets \varnothing$\;
$\mathcal{A}^{\prime}_{\mathrm{sel}} \gets \varnothing$\;

\While{\textup{$\exists$ unscheduled tasks of candidate nodes in $\mathit{\Lambda}$}}{
    $applCandNodeExists \gets 0$\;
    $minEFT \gets 0$\;
    
    Select first unscheduled task $\tau_i$ (i.e., $N_i^{\prime} \in G^{\prime}$) according to $\mathit{\Lambda}$\;
    
    \ForEach{\textup{candidate node $N_{i \lambda k.q} \in N_i^{\prime}$}}{

        $\mathcal{N}^{\prime}_{\mathrm{temp}} \gets \mathcal{N}^{\prime}_{\mathrm{sel}} \cup \{ N_{i \lambda k.q} \}$\;

        $\mathcal{A}^{\prime}_{\mathrm{temp}} \gets \mathcal{A}^{\prime}_{\mathrm{sel}} \cup \{ A_{j \mu l.r \rightarrow i\lambda k.q} \in \mathcal{A}^{\prime} | N_{j \mu l.r} \in \mathcal{N}^{\prime}_{\mathrm{sel}}\}$\;

        $energyBgtExceeded \gets 0$\;
        \ForEach{\textup{device $u_{\mu l} \in \mathcal{U}$}}{
       
            \If{\textup{constraint \eqref{eq:fourteen} is violated for $u_{\mu l}$ using $\mathcal{N}^{\prime}_{\mathrm{temp}}, \mathcal{A}^{\prime}_{\mathrm{temp}}$ in place of $\mathcal{N}^{\prime}, \mathcal{A}^{\prime}$, respectively}}{
                $energyBgtExceeded \gets 1$\;
                Break\;           
            }
        }
        \eIf{\textup{$energyBgtExceeded \lor M_i > M_{\lambda k}^{\mathrm{bgt}} \lor S_i > S_{\lambda k}^{\mathrm{bgt}}$}}{Continue\tcp*[r]{$N_{i \lambda k.q}$ is not applicable}}
        {
            
            $t_i \gets \hspace{-6pt} \max\limits_{A_{j \mu l.r \rightarrow i \lambda k.q} \in \mathcal{A}^{\prime}_{\mathrm{temp}}} \hspace{-8pt} \{ t_j +L_{j \mu l.r} + CL_{j \mu l.r \rightarrow i \lambda k.q}\}$\;
        
            $\mathit{\Lambda}_{\lambda k} \gets \{ N_{j \lambda k.r} \in \mathcal{N}^{\prime}_{\mathrm{sel}} | t_j + L_{j \lambda k.r} > t_i \}$\;
            
            Sort list $\mathit{\Lambda}_{\lambda k}$ by non-decreasing order of $t_j + L_{j \lambda k.r}$\;

            \For{\textup{$n \gets 1$ \KwTo $|\mathit{\Lambda}_{\lambda k}|$}}{
                
                $N_{j \lambda k.r} \gets \mathit{\Lambda}_{\lambda k}[n]$\;
                
                $\mathcal{N}^{\prime}_{\mathrm{temp}} \gets \{N_{i \lambda k.q}\} \cup \{ N_{h \lambda k.s} \in \mathit{\Lambda}_{\lambda k} | s \neq q, t_i < (t_h + L_{h \lambda k.s}) \land (t_i + L_{i \lambda k.q}) > t_h \}$\;
                
                \If{\textup{ $(r \hspace{-1pt} = \hspace{-1pt} q \hspace{-1pt} \land \hspace{-1pt} t_i \hspace{-1pt} < \hspace{-1pt} (t_j \hspace{-1pt} + \hspace{-1pt} L_{j \lambda k.r}) \hspace{-1pt} \land \hspace{-1pt} (t_i \hspace{-1pt} + \hspace{-1pt} L_{i \lambda k.q}) \hspace{-1pt} > \hspace{-1pt} t_j)
                \allowbreak \lor \hspace{-2pt} \bigg(\sum\limits_{N_{h \lambda k.s} \in \mathcal{N}^{\prime}_{\mathrm{temp}}} \hspace{-15pt} y_{\lambda k}^{c_a}\,z_h^{c_a} > 1 \land z_i^{c_a}=1 \land a>0 \bigg) \allowbreak 
                \lor \hspace{-5pt} \sum\limits_{N_{h \lambda k.s} \in \mathcal{N}^{\prime}_{\mathrm{temp}}} \hspace{-15pt} M_h > M_{\lambda k}^{\mathrm{bgt}}
                \allowbreak \, \lor \hspace{-5pt} \sum\limits_{N_{h \lambda k.s} \in \mathcal{N}^{\prime}_{\mathrm{temp}}} \hspace{-15pt} S_h > S_{\lambda k}^{\mathrm{bgt}}$ 
                }}{
                    $t_i \gets t_j + L_{j \lambda k.r}$\;
                }

                $n \gets n + 1$\;
            }

            $EFT_{i \lambda k.q} \gets t_i + L_{i \lambda k.q}$\;
            
            \If{\textup{$minEFT = 0 \lor EFT_{i \lambda k.q} < minEFT$}}{
                $minEFT \gets EFT_{i \lambda k.q}$\;
            }
                
            $applCandNodeExists \gets 1$\;
        }
    }

    \eIf{$applCandNodeExists \land minEFT \leq L_{\mathrm{thr}}$}{
        Select candidate node $N_{i \lambda k.q}$ that has $EFT_{i \lambda k.q} = minEFT$\;
        $x_{i \lambda k.q} \gets 1$\; 
        Mark task $\tau_i$ as scheduled\;
        $\mathcal{N}^{\prime}_{\mathrm{sel}} \gets \mathcal{N}^{\prime}_{\mathrm{sel}} \cup \{ N_{i \lambda k.q}\}$\;
        \ForEach{$A_{j \mu l.r \rightarrow i \lambda k.q} \in \mathcal{A}^{\prime}$}{
            \If{$N_{j \mu l.r} \in \mathcal{N}^{\prime}_{\mathrm{sel}}$}{
                $x_{j \mu l.r \rightarrow i \lambda k.q} \gets 1$\;
                $\mathcal{A}^{\prime}_{\mathrm{sel}} \gets \mathcal{A}^{\prime}_{\mathrm{sel}} \cup \{ A_{j \mu l.r \rightarrow i \lambda k.q} \}$\;
            }
        }
    }
    {
        $isInfeasible \gets 1$\; 
        Break\;
    }    
}
\Return{$isInfeasible$}\;

\caption{Enhanced version of HEFT.}\label{alg:heft}

\end{algorithm}

In the first phase, the upward rank of a candidate node $N_{i \lambda k.q}$ (line 2) is calculated based on the upward rank of its child nodes \cite{Topcuoglu2002}:
\begin{equation}
\label{eq:rank}
R_{i \lambda k\hspace{-0.3pt}.\hspace{-0.3pt}q} \hspace{-2pt} = \hspace{-2pt} L_{i \lambda k\hspace{-0.3pt}.\hspace{-0.3pt}q} \hspace{0.75pt} + \hspace{-2pt} \max_{A_{i \lambda k\hspace{-0.3pt}.\hspace{-0.3pt}q \rightarrow \hspace{-0.3pt} j \mu l\hspace{-0.3pt}.\hspace{-0.3pt}r} \in \mathcal{A}^{\prime}} \hspace{-2pt} \{ \hspace{-0.75pt} CL_{i \lambda k\hspace{-0.3pt}.\hspace{-0.3pt}q \rightarrow \hspace{-0.3pt} j \mu l\hspace{-0.3pt}.\hspace{-0.3pt}r} \hspace{0.25pt} + \hspace{0.75pt} R_{j \mu l\hspace{-0.3pt}.\hspace{-0.3pt}r} \hspace{-0.75pt} \}\hspace{-1pt}.
\end{equation}
Candidate nodes are prioritized based on their upward rank (line 5).
In the second phase, for each unscheduled task $\tau_i$ (for which no candidate node has been selected) we examine each of its candidate nodes $N_{i \lambda k.q} \in N_i^{\prime}$ (in order of rank) to select the one that minimizes its finish time.
In particular, we first check if selecting a candidate node $N_{i \lambda k.q}$ would exceed the energy budget of any device (lines 17--23). Moreover, we examine whether the memory or storage requirements of $\tau_i$ exceed the respective budgets of device $u_{\lambda k}$ (line 23).
If any of these conditions hold, we skip $N_{i \lambda k.q}$ and continue with the next (according to its rank) candidate node of $\tau_i$ (line 24). 

Otherwise, using $N_{i \lambda k.q}$ we determine the start time $t_i$ that yields the earliest finish time $EFT_{i \lambda k.q}$ for $\tau_i$ on processing core $p_{\lambda k.q}$, so that its precedence constraints are satisfied (line 26), its execution does not overlap with other tasks already allocated on $p_{\lambda k.q}$, and the capability, memory, and storage capacities of device $u_{\lambda k}$ are not exceeded (lines 27-37).
For the capacity constraints, we consider tasks already allocated on $u_{\lambda k}$ (but on different processing cores) and whose execution will overlap with $\tau_i$ (line 31).
Subsequently, we select the candidate node that minimizes the earliest finish time of $\tau_i$ on any device/processing core, without exceeding the execution deadline $L_{\mathrm{thr}}$ (lines 44--54). If there is a task for which no candidate node can satisfy the above constraints, then the problem is infeasible.
Considering that for dense ETGs the number of arcs $|\mathcal{A}^{\prime}|$ is proportional to $|\mathcal{N}^{\prime}|^2$ \cite{Topcuoglu2002}, the worst-case time complexity of the first phase is dominated by the operations performed in lines 1--3. Thus, its worst-case time complexity is $O(|\mathcal{N}^{\prime}|^3)$.
Similarly, the worst-case time complexity of the second phase is dominated by the operations in lines 17--22, i.e., $O(|\mathcal{U}| |\mathcal{N}^{\prime}|^3)$. Evidently, the time complexity of the second phase is greater than that of the first phase. Hence, the overall worst-case time complexity of extended HEFT is $O(|\mathcal{U}||\mathcal{N}^{\prime}|^3)$.     
It is noted that the time complexity of the proposed optimal MILP approach depends not only on the size of the problem, but also on the utilized solver. Commercial solvers (e.g., Gurobi \cite{gurobi}) typically employ proprietary algorithms whose implementation details are not publicly available, and thus their time complexity cannot be easily derived \cite{Mo2023}.
With regard to problem size, as mentioned in \cref{subsec:etg}, in the worst case the number of nodes and arcs in the resulting ETG\,$G^{\prime}$ increases linearly and quadratically, respectively, compared to TG\,$G$.  

\section{Experimental Evaluation}
\label{sec:evaluation}
We evaluated and compared the proposed MILP approach against the enhanced version of HEFT, considering a real-world workflow under various system configurations. 
To further validate and examine the scalability of our technique to applications of different sizes, we developed and used synthetic workflows with appropriate parameters.

\setlength{\textfloatsep}{\oldtextfloatsep}

\begin{table}[t]
    \centering
    \caption{Device Capabilities}
    \vspace{-5pt}
    \scriptsize
    \resizebox{0.75\columnwidth}{!}{
        \begin{tabular}{ll}
            \toprule
            \# & Description\\
            \hline
            0 & Basic computational capability\\
            1 & Thermal camera\\
            2 & LiDAR sensor\\
            3 & Multispectral camera\\
            4 & High-precision GNSS module\\
            5 & Custom payload release mechanism\\
            6 & Specialized software module (e.g., for UAV coordination)\\
            7 & Integrated display unit\\
            8 & Specialized hardware accelerator (e.g., a GPU)\\
            9 & High-availability storage\\
            \bottomrule
        \end{tabular}
    }
    \label{tab:capabilities}
    \vspace{-10pt}
\end{table}

\vspace{-5pt}
\subsection{Experimental Setup}
\label{subsec:setup}
We investigated an edge-hub-cloud CPS with multiple edge devices in six different configurations (C1--C6). In each configuration, we considered varying numbers and types of edge devices that could communicate with each other and a hub device, which in turn could communicate with a cloud server. 
The devices were based on typical real-world counterparts, featuring heterogeneous multicore processors, and various memory, storage, and energy capacities.
All devices featured different sensing/actuating or other specialized capabilities based on the real-world use case. These capabilities are indicated by integers in the range $[0,9]$, as shown in \cref{tab:capabilities}.

The number of reserved processing cores $\gamma_{\lambda k}$ and the budgets $M_{\lambda k}^{\mathrm{bgt}}$, $S_{\lambda k}^{\mathrm{bgt}}$, and $E_{\lambda k}^{\mathrm{bgt}}$ were a subset of the respective ones featured by each device, as computational resources in such use cases are typically limited and shared among different applications.
\cref{tab:devices} shows the system devices and their specifications, the considered budgets, as well as the capabilities of each device (per configuration).
In addition to the hub device (Mi Notebook Pro) and the cloud server (HPE DL580 Gen10), each configuration included two (C1, C2), three (C3, C4), or four (C5, C6) edge devices selected from Raspberry Pi 3, Odroid XU4, Jetson TX2, and Jetson Xavier NX. Each edge device was considered to be attached to a UAV with different sensing/actuating capabilities. 
For example, configuration C1 included Raspberry Pi 3 with capabilities $\{0, 1, 3\}$, Jetson Xavier NX with capabilities $\{0, 2, 4, 5\}$, Mi Notebook Pro with capabilities $\{0, 6, 7\}$, and HPE DL580 Gen10 with capabilities $\{0, 8, 9\}$, as shown in the respective column of \cref{tab:devices}.
\cref{tab:channels} includes the ranges of the bandwidth and energy parameters for the communication channels between each pair of devices, which were based on real-world measurements \cite{Huang2012, Vladan2021}. The model of the CPS under study is illustrated in \cref{fig:system}.
We implemented both the proposed MILP method and the extended version of HEFT in C++.  
In the MILP approach, the formulated problem was solved using Gurobi Optimizer 11 \cite{gurobi}, on a server equipped with an Intel Xeon Gold 6240 processor\,@\,2.6\,GHz and 400\,GiB of RAM.

\begin{table*}[!t]
    \setlength{\tabcolsep}{1.5pt}
    \centering
    \caption{System Devices \& Configurations}
    \vspace{-5pt}
    \resizebox{\textwidth}{!}{
    \begin{tabular}{
    @{\extracolsep{1.5pt}}
    cccrrcrccrrllllll}
    \toprule
    \multirow{3}{*}{$u_{\lambda k}$} & \multirow{3}{*}{Device} & \multicolumn{4}{c}{Specifications} & \multicolumn{1}{c}{\multirow{2}{*}{Perf.}} & \multirow{3}{*}{$\gamma_{\lambda k}$} & \multicolumn{3}{c}{Budgets} & \multicolumn{6}{c}{Capabilities per Configuration${}^{4}$}\\
    \cline{3-6}
    \cline{9-11}
    \cline{12-17}
                    &        & \multirow{2}{*}{Processor} & \multicolumn{1}{c}{Memory} & \multicolumn{1}{c}{Storage} & Battery${}^{3}$ & \multicolumn{1}{c}{\multirow{2}{*}{Ratio}} &  & $M_{\lambda k}^{\mathrm{bgt}}$ & \multicolumn{1}{c}{$S_{\lambda k}^{\mathrm{bgt}}$} & \multicolumn{1}{c}{$E_{\lambda k}^{\mathrm{bgt}}$} & \multirow{2}{*}{\hspace{5pt} C1} & \multirow{2}{*}{\hspace{5pt} C2} & \multirow{2}{*}{\hspace{5pt} C3} & \multirow{2}{*}{\hspace{5pt} C4} & \multirow{2}{*}{\hspace{5pt} C5} & \multirow{2}{*}{\hspace{5pt} C6}\\
                    & & & \multicolumn{1}{c}{(GiB)} & \multicolumn{1}{c}{(GiB)}  & (Wh) & & & (GiB) & \multicolumn{1}{c}{(GiB)} & \multicolumn{1}{c}{(Wh)} & & & & & & \\
    \hline

    $u_{\mathrm{e}1}$ & Raspberry\,Pi\,3   & Cortex-A53\,@\,1.4\,GHz                              & 1 \hspace{5pt}   & 16 \hspace{5pt}    & 33.3 & 1.00  & 2 & 0.95 & 1.0 \hspace{3pt}  & 1 \hspace{3pt}  & $\{0, 1, 3\}$    & \,$\varnothing$  & $\{0, 3\}$       & $\{0, 3\}$       & $\{0, 1\}$    & $\{0, 1\}$\\
    $u_{\mathrm{e}2}$ & Odroid\,XU4        & Cortex-A7\,\&\,Cortex-A15${}^{1}$\,@\,2.0\,GHz       & 2 \hspace{5pt}   & 16 \hspace{5pt}    & 33.3 & 1.20  & 2 & 1.00 & 1.5 \hspace{3pt}  & 1 \hspace{3pt}  & \,$\varnothing$  & $\{0, 1, 3\}$    & $\{0, 5\}$       & $\{0, 5\}$       & $\{0, 3\}$    & $\{0, 3\}$ \\
    $u_{\mathrm{e}3}$ & Jetson\,TX2        & NVIDIA\,Denver2\,\&\,Cortex-A57${}^{2}$\,@\,2.0\,GHz & 8 \hspace{5pt}   & 32 \hspace{5pt}    & 33.3 & 2.80  & 2 & 2.00 & 2.0 \hspace{3pt}  & 1 \hspace{3pt}  & \,$\varnothing$  & $\{0, 2, 4, 5\}$ & \,$\varnothing$  & $\{0, 1, 2, 4\}$ & $\{0, 5\}$    & $\{0, 2, 4\}$ \\ 
    $u_{\mathrm{e}4}$ & Jetson\,Xavier\,NX & NVIDIA\,Carmel ARMv8.2\,@\,1.4\,GHz                  & 8 \hspace{5pt}   & 32 \hspace{5pt}    & 33.3 & 5.74  & 2 & 2.00 & 2.5 \hspace{3pt}  & 1 \hspace{3pt}  & $\{0, 2, 4, 5\}$ & \,$\varnothing$  & $\{0, 1, 2, 4\}$ & \,$\varnothing$  & $\{0, 2, 4\}$ & $\{0, 5\}$ \\
    $u_{\mathrm{h}1}$ & Mi\,Notebook\,Pro  & Intel\,i5\,8250U\,@\,1.6\,GHz                        & 8 \hspace{5pt}   & 512 \hspace{5pt}   & 60.0 & 15.23 & 4 & 3.00 & 5.0 \hspace{3pt}  & 2 \hspace{3pt}  & $\{0, 6, 7\}$    & $\{0, 6, 7\}$    & $\{0, 6, 7\}$    & $\{0, 6, 7\}$    & $\{0, 6, 7\}$ & $\{0, 6, 7\}$ \\
    $u_{\mathrm{c}1}$ & HPE\,DL580\,Gen10  & Intel\,Xeon\,Gold\,6240\,@\,2.6\,GHz                 & 400 \hspace{5pt} & 10240 \hspace{5pt} & --   & 21.70 & 6 & 4.00 & 10.0 \hspace{3pt} & 10 \hspace{3pt} & $\{0, 8, 9\}$    & $\{0, 8, 9\}$    & $\{0, 8, 9\}$    & $\{0, 8, 9\}$    & $\{0, 8, 9\}$ & $\{0, 8, 9\}$ \\
    \bottomrule
    
    \multicolumn{17}{l}{${}^{1,2}$Without loss of generality, the reserved processing cores are considered to be located on Cortex-A15 and Cortex-A57, respectively.}\\
    \multicolumn{17}{l}{${}^{3}$A compatible external battery (TalentCell YB1203000-USB) is considered for the edge devices. \quad ${}^{4}$$\varnothing$ denotes that a device is not included in a configuration.}
    \end{tabular}
    }
    \label{tab:devices}
    \vspace{-25pt}
\end{table*}

\vspace{-5pt}
\subsection{Real-World Workflow}
\label{subsec:real}

\subsubsection{Overview}
\label{subsubsec:realOverview}
\begin{figure}[t]
    \centering
    \begin{minipage}[b]{0.63\columnwidth}
        \begin{table}[H]
        \centering
        \caption{Real-World Workflow Tasks}
        \vspace{-5pt}
        \resizebox{\columnwidth}{!}{
            \begin{tabular}{ll}
                \toprule
                Task & Description\\
                \hline
                $N_1$ & Capture multispectral image\\
                $N_2$ & Multispectral image preprocessing\\
                $N_3$ & Detect power transmission lines\\
                $N_4$ & Detect vegetation encroachment\\
                $N_5$ & Perform LiDAR scan\\
                $N_6$ & Data preprocessing/integrate GNSS data\\
                $N_7$ & Detect power towers\\
                $N_8$ & Detect structural integrity problems\\
                $N_9$ & Capture thermal image\\
                $N_{10}$ & Thermal image preprocessing\\
                $N_{11}$ & Detect overheating components\\
                $N_{12}$ & Multi-source data fusion\\
                $N_{13}$ & Save fused data on high-availability storage\\
                $N_{14}$ & Path planning/coordinate tag deployment\\
                $N_{15}$ & Display final output\\
                $N_{16}$ & Deploy tags at problematic sections\\
                \bottomrule
            \end{tabular}
        }
        \label{tab:realTasks}
        \end{table}
    \end{minipage}   
    \hfill
     \begin{minipage}[b]{0.35\columnwidth}
        \centering
        \includegraphics[width=0.95\columnwidth]{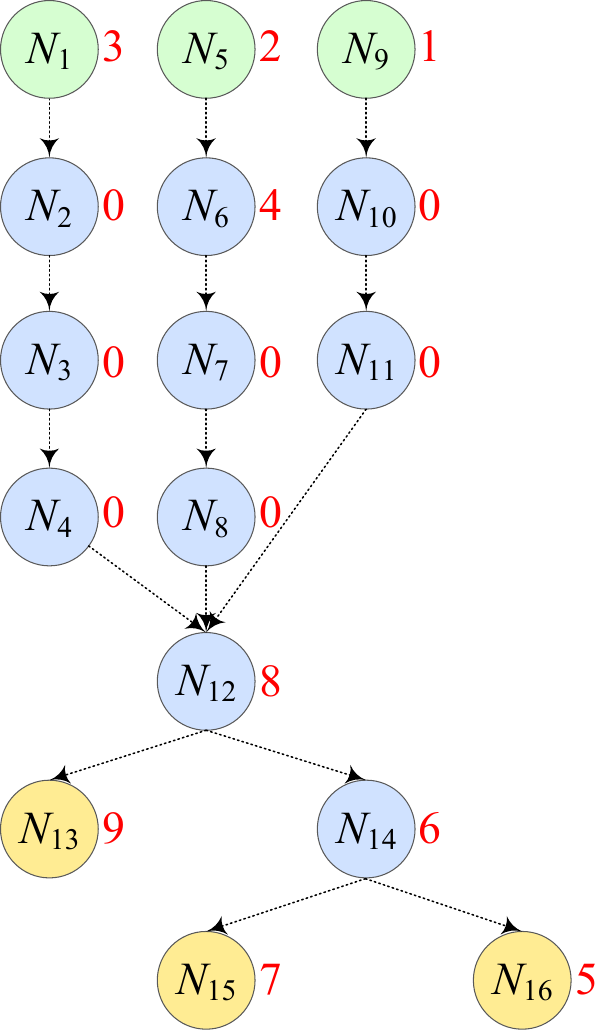}
        \vspace{-5pt}
        \caption{Real-world workflow TG.}
        \label{fig:realWorkflow}
    \end{minipage}
    \vspace{-12pt}
\end{figure}

We considered a relevant real-world workflow for the autonomous UAV-based inspection of power transmission towers and lines, based on \cite{Savva2021}.
It consists of 16 tasks, as shown in \cref{tab:realTasks} and \cref{fig:realWorkflow}. 
Entry and exit tasks are depicted in green and yellow, respectively, while intermediate tasks are shown in blue. The required capabilities of the tasks (as described in \cref{tab:capabilities}) are indicated by the red integers (0--9).
In this use case, multiple UAVs with attached edge devices collaborate to capture multispectral images to detect power transmission lines and vegetation encroachment (tasks $N_1$--$N_4$), perform LiDAR scans and integrate them with high-precision global navigation satellite system (GNSS) data to detect power towers and structural integrity problems (tasks $N_5$--$N_8$), and capture thermal images to detect overheating components such as insulators (tasks $N_9$--$N_{11}$).

The diverse data from all UAVs are fused to create a comprehensive visual representation of the infrastructure (task $N_{12}$).
This is a computationally intensive process necessitating a specialized hardware accelerator, such as a high-performance GPU provided by a cloud server \cite{Gao2022}. The fused data are saved on high-availability storage in the cloud to ensure enhanced reliability and facilitate remote access (task $N_{13}$). In addition, the fused data are used for tag deployment path planning and coordination, a process requiring a specialized software module for UAV control, typically installed on the hub device to improve latency (task $N_{14}$). Finally, the output is displayed on the hub device (task $N_{15}$), while a UAV starts deploying location-transmitting tags at the identified problematic sections of the infrastructure (to facilitate the ground crew in locating them) using a custom payload release mechanism (task $N_{16}$).

For the execution of the application, we considered two, three, or four edge devices (each attached to a different UAV with varied capabilities), as well as a hub device and a cloud server, based on the configurations C1--C6 in \cref{tab:devices}.
We transformed the TG of the application into the corresponding ETGs, one for each system configuration. 
The number of ETG candidate nodes/arcs in each case is shown in \cref{tab:realETGs}.
The ETG parameters $D_i$, $M_i$, $S_i$, $L_{i \lambda k.q}$, and $P_{i \lambda k.q}$ were determined using profiling and power monitoring tools (perf and Powertop) \cite{powertop} or through relevant benchmarks (Phoronix Test Suite) \cite{openbenchmarking}. The ranges of these parameters are listed in \cref{tab:realParams}. $E_{i \lambda k.q}$, $CL_{i \lambda k.q \rightarrow j \mu l.r}$, and $CE_{i \lambda k.q \rightarrow j \mu l.r}$ were calculated using \eqref{eq:compEnergy}, \eqref{eq:commLatency}, and \eqref{eq:commEnergy}, respectively, whereas $Q_i$ and $\sigma_{i \lambda k.q \rightarrow j \mu l.r}^{\nu m}$ were derived from the respective TG and ETG structures. For each ETG, we set the deadline $L_{\mathrm{thr}}$ to be 1.5 times greater than its critical path (considering the computational and communication latency of the candidate nodes and arcs, respectively), as this was a realistic, but also challenging scenario \cite{Bai2021}.

\subsubsection{Evaluation Results}
\label{subsubsec:realResults}
\begin{figure*}[!t]
    \centering
    \subfloat[]{\includegraphics[width=.29\textwidth]{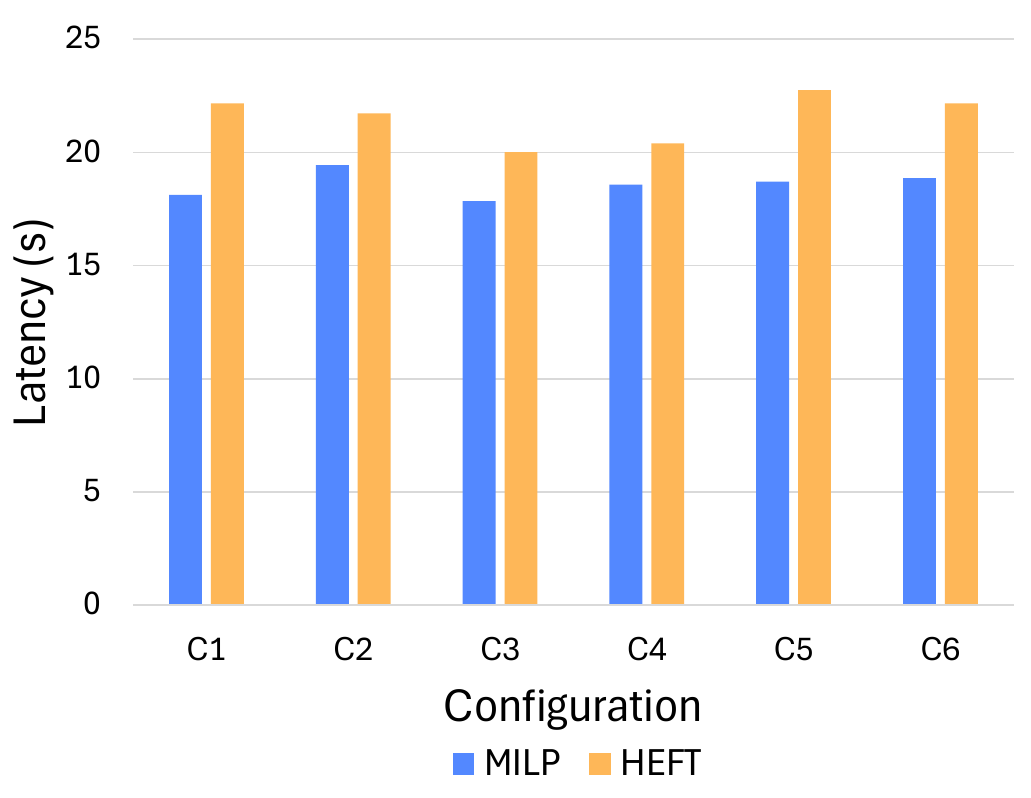}
        \label{fig:realLatency}}
    \hfill
    \subfloat[]{\includegraphics[width=.29\textwidth]{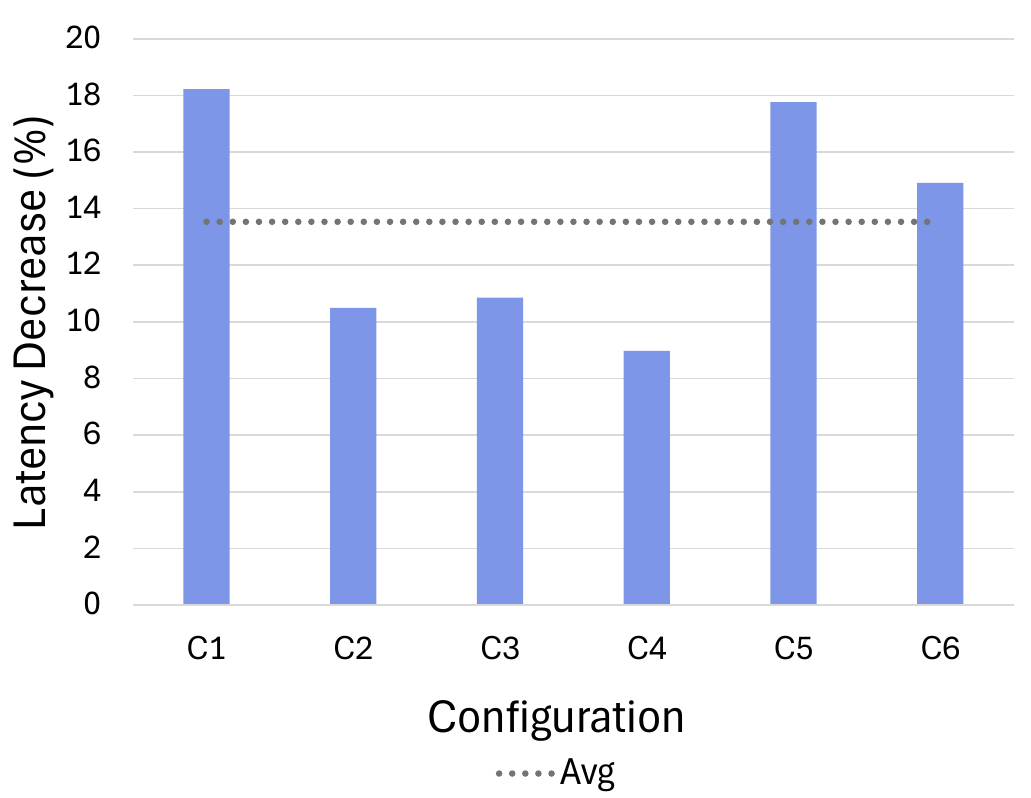}
        \label{fig:realLatencyDecrease}}
    \hfill
    \subfloat[]{\includegraphics[width=.29\textwidth]{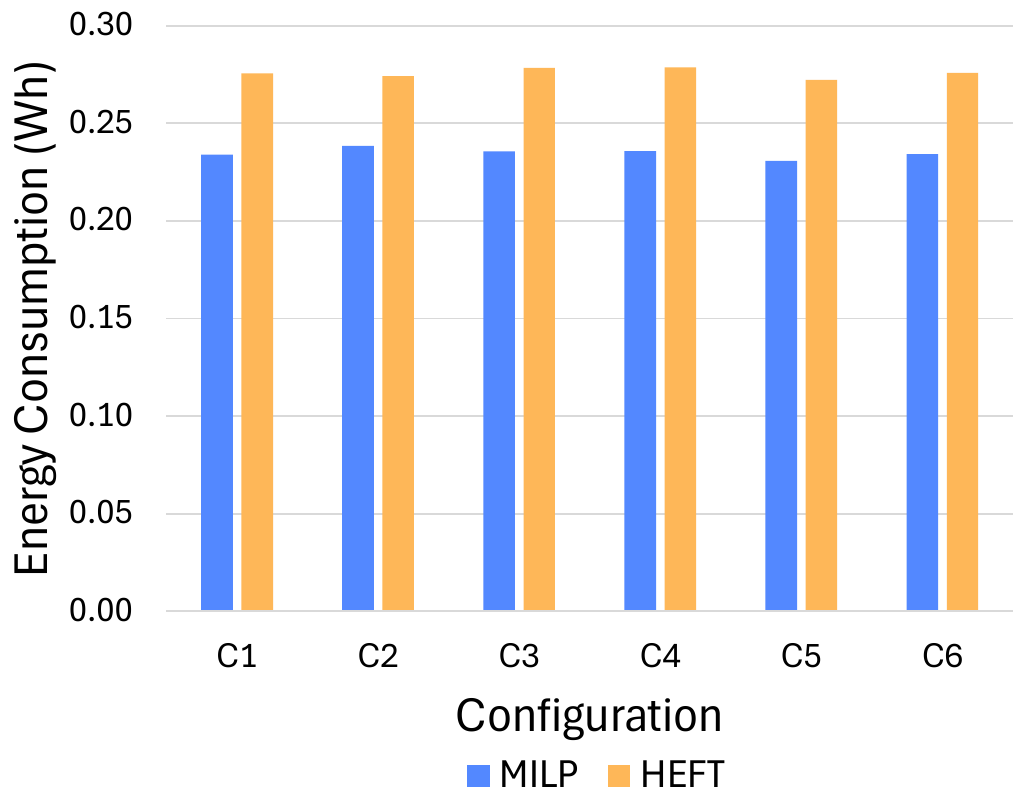}
        \label{fig:realEnergy}}
    \vspace{-3pt}
    \caption{Comparative evaluation between proposed MILP approach and enhanced version of HEFT for the real-world workflow under all system configurations.}
    \label{fig:realResults}
    \vspace{-12pt}
\end{figure*}

We used the real-world workflow to compare our MILP approach with the extended version of HEFT.
Our evaluation was primarily focused on overall latency, which is our problem objective \eqref{eq:objective}. In addition, we considered the overall energy consumption as a secondary metric. It was calculated based on the resulting schedule in each case, by summing the left-hand side of \eqref{eq:fourteen} for all system devices.
\cref{fig:realLatency} showcases the results of the comparative evaluation between MILP and HEFT for each system configuration, with respect to overall latency. It can be observed that MILP consistently outperformed HEFT in all of the examined scenarios.
This superiority is further highlighted in \cref{fig:realLatencyDecrease}, which demonstrates the latency improvement achieved by MILP over HEFT, under each configuration. Notably, the proposed MILP technique provided an average latency decrease of 13.54\%, across all cases.

\begin{figure}[t]
    \centering
    \begin{minipage}[b]{0.48\columnwidth}
        \includegraphics[width=\columnwidth]{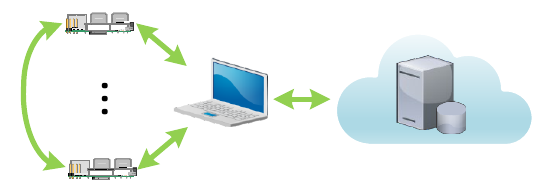}
        \caption{System model.}
        \label{fig:system}
    \end{minipage}
    \hfill
    \begin{minipage}[b]{0.5\columnwidth}
        \begin{table}[H]
        \setlength{\tabcolsep}{1.5pt}
        \centering
        \caption{Communication Channels}
        \vspace{-5pt}
        \resizebox{\columnwidth}{!}{
        \begin{tabular}{lccc}
        \toprule
        \multicolumn{1}{c}{Comm.} & $\theta_{\lambda k, \mu l}$ & $\pi_{\lambda k, \mu l}$ & $\rho_{\lambda k, \mu l}$\\
        \multicolumn{1}{c}{Channel} & (Mbit/s) & (\SI{}{\micro \joule}/bit) & (\SI{}{\micro \joule}/bit)\\
        \hline
        \hspace{3pt} $u_{\mathrm{e} k} \leftrightarrow u_{\mathrm{e} l}$ & $[6,9]$ & $[0.6,1.0]$ & $[0.4,0.6]$\\
        \hspace{3pt} $u_{\mathrm{e} k} \rightarrow u_{\mathrm{h} 1}$ & $[9,13]$ & $[0.8,1.2]$ & $[0.6,0.8]$\\
        \hspace{3pt} $u_{\mathrm{h} 1} \rightarrow u_{\mathrm{e} l}$ & $[7,10]$ & $[0.7,1.1]$ & $[0.5,0.7]$\\
        \hspace{3pt} $u_{\mathrm{h} 1} \rightarrow u_{\mathrm{c} 1}$ & $[10,15]$ & $[1.8,2.7]$ & $[0.8,1.2]$\\
        \hspace{3pt} $u_{\mathrm{c} 1} \rightarrow u_{\mathrm{h} 1}$ & $[16,24]$ & $[2.0,3.0]$ & $[1.0,1.5]$\\
        \bottomrule
        \end{tabular}
        }
        \label{tab:channels}
        \end{table}
    \end{minipage}
    \vspace{-25pt}
\end{figure}

\cref{fig:realLatency,fig:realLatencyDecrease} also show that the overall latency yielded by each method, as well as the latency improvement attained by MILP, depended on the employed system configuration. In particular, the latency decrease presented significant fluctuations, ranging from 8.98\% to 18.22\% across all configurations.
This indicates that in the examined problem, the number, type, and sensing/actuating capabilities of edge devices play a decisive role in the resulting overall latency.
The poor performance of HEFT was due to its inherent limitation in making sequential scheduling decisions. In contrast to MILP, HEFT schedules one task at a time, without assessing the entire solution space. Consequently, in HEFT a decision that is optimal at a specific step does not necessarily lead to a globally optimal schedule.

Regarding the overall energy consumption, which served as a secondary metric, \cref{fig:realEnergy} shows that MILP yielded a lower consumption than HEFT in all scenarios, even though the minimization of energy was not the objective of either technique.
Specifically, MILP provided average energy savings of 14.88\% over HEFT. 
The time required by the Gurobi solver to return a solution for the proposed MILP approach ranged between 6.68 and 20.62\,s, as shown in \cref{tab:realETGs}.
\cref{tab:realETGs} also demonstrates the number of generated variables and constraints in each case. 
Considering the pre-programmed nature of the examined workflow, which allows for its scheduling to be performed offline, the solver runtime is short and practical.

\begin{figure}[t]
    \vspace{-5pt}
    \centering   
    \begin{minipage}[b]{0.58\columnwidth}
        \begin{table}[H]
        \setlength{\tabcolsep}{1.5pt}
        \centering
        \caption{Real-World Workflow ETGs}
        \vspace{-5pt}
        \resizebox{0.9\columnwidth}{!}{
            \begin{tabular}{cccr}
                \toprule
                \multirow{2}{*}{Config.} & \multirow{2}{*}{\#Nodes/Arcs} & \#Variables/ & \multicolumn{1}{c}{Solver}\\
                 & & Constraints & \multicolumn{1}{c}{Runtime (s)}\\
                \hline
                C1 & 128\,/\,1208 & 5705\,/\,16353 & 6.68\hspace{10pt} \\
                C2 & 128\,/\,1208 & 5705\,/\,16353 & 8.71\hspace{10pt}  \\
                C3 & 142\,/\,1496 & 6455\,/\,18736 & 11.19\hspace{10pt} \\
                C4 & 142\,/\,1496 & 6455\,/\,18736 & 13.05\hspace{10pt} \\
                C5 & 156\,/\,1816 & 7237\,/\,21243 & 14.80\hspace{10pt} \\
                C6 & 156\,/\,1816 & 7237\,/\,21243 & 20.62\hspace{10pt} \\
                \bottomrule
            \end{tabular}
        }
        \label{tab:realETGs}
        \end{table}
    \end{minipage}
    \hfill
    \begin{minipage}[b]{0.4\columnwidth}
        \begin{table}[H]
        \setlength{\tabcolsep}{1.5pt}
        \centering
        \caption{Real-World Workflow\\ETG Parameter Ranges}
        \vspace{-5pt}
        \resizebox{0.9\columnwidth}{!}{
            \begin{tabular}{lc}
                \toprule
                Param. & Range\\
                \hline
                $D_i$ & $[3.22, 151.27]$\,Mbit\\
                $M_i$ & $[103.98, 3800.17]$\,Mbit\\
                $S_i$ & $[246.57, 3766.42]$\,Mbit\\
                $L_{i \lambda k.q}$ & $[2.57, 12648.38]$\,ms\\
                $P_{i \lambda k.q}$ & $[0.30, 23.70]$\,W\\
                \bottomrule
            \end{tabular}
        }
        \label{tab:realParams}
        \end{table}
    \end{minipage}
    \vspace{-20pt}
\end{figure}

\vspace{-5pt}
\subsection{Scalability Analysis}
\label{subsec:scalability}

\subsubsection{Generation of Synthetic Workflows}
\label{subsubsec:syntheticWorkflows}
To investigate the scalability of the proposed MILP approach to workflows of various sizes, we generated random TGs using the generator in \cite{Dick1998}. 
Specifically, based on the real-world workflow in \cref{subsec:real}, and considering that similar cyber-physical applications for the examined system architecture typically have a coarse-grained structure with a small-to-moderate number of tasks \cite{Zheng2021, Alam2017, Kashino2019, Savva2021}, we generated 25 random TGs grouped into five sets of different sizes. Each set comprised five TGs with 10,\hspace{2pt}20,\hspace{2pt}30,\hspace{2pt}40, or 50 tasks/nodes, and an average in/out degree (incoming/outgoing arcs per node) of\hspace{2pt}1.7/1.7. We randomly assigned specialized capabilities to the tasks of each TG, considering an 80\% probability for the entry and exit tasks (as they usually require a specific sensor or actuator, respectively), and a 20\% probability for the intermediate tasks. The remaining tasks were assigned the basic computational capability.

We subsequently transformed the generated TGs into the corresponding ETGs, based on configuration C6 (shown in \cref{tab:devices}).
We selected the particular configuration as it was among the most complex ones, encompassing four edge devices. 
The average number of nodes/arcs in each TG and ETG is shown in \cref{tab:syntheticETGs}.
For the ETG candidate node parameters $D_i$, $M_i$, and $S_i$ (which are device-independent) we randomly assigned values from the ones obtained in the real-world use case. 
For $L_{i \lambda k.q}$ and $P_{i \lambda k.q}$ (which are device/core-dependent), we first assigned values from those measured on Raspberry Pi 3 (used as a reference device) in the real-world scenario.
Regarding the other devices, $L_{i \lambda k.q}$ and $P_{i \lambda k.q}$ were calculated based on the performance ratio of each device with respect to the reference device.
The performance ratio (shown in \cref{tab:devices}) was defined by comparing the performance score of each device to that of the reference device. The performance scores were obtained by running the Phoronix Test Suite benchmarks \cite{openbenchmarking} on all devices (including the reference device). The remaining ETG candidate node and arc parameters were determined as described in \cref{subsubsec:realOverview}.
Our synthetic workflow datasets are openly accessible at \cite{Zenodo_COMPSAC}.

\begin{table}[t]
    \centering
    \caption{Synthetic Workflow TGs \& ETGs}
    \vspace{-5pt}
    \scriptsize
    \resizebox{0.85\columnwidth}{!}{
        \begin{tabular}{ccccr}
            \toprule
            TG Size   & TG Avg.  & ETG Avg.      & Avg. \#Variables/ & \multicolumn{1}{c}{Avg. Solver}\\
            (\#Nodes) & \#Arcs   & \#Nodes/Arcs  & Constraints       & \multicolumn{1}{c}{Runtime (min)}\\
            \hline
            10 & 17 & 131\,/\,2871 & 5729\,/\,18846 & 0.20\hspace{15pt} \\
            20 & 36 & 240\,/\,6275 & 16553\,/\,52108 & 3.80\hspace{15pt} \\
            30 & 45 & 388\,/\,8464 & 33989\,/\,102391 & 31.78\hspace{15pt} \\
            40 & 65 & 492\,/\,11455 & 52916\,/\,157228 & 43.65\hspace{15pt} \\
            50 & 89 & 621\,/\,16917 & 82169\,/\,243625 & 99.24\hspace{15pt} \\
            \bottomrule
        \end{tabular}
    }
    \label{tab:syntheticETGs}
    \vspace{-12pt}
\end{table}

\subsubsection{Scalability Results}
\label{subsubsec:syntheticResults}
\begin{figure}[!t]
    \centering
    \includegraphics[width=0.65\columnwidth]{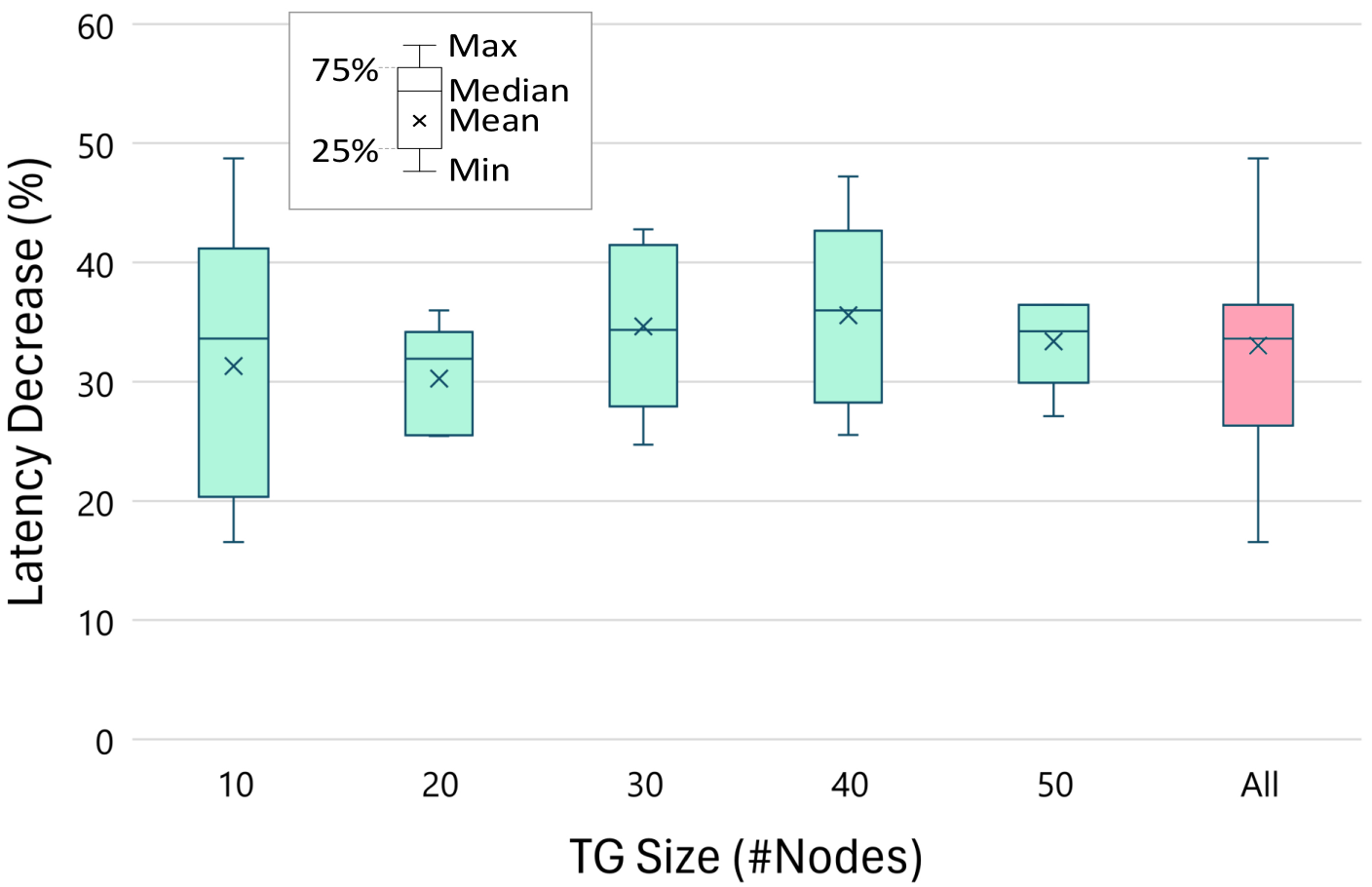}
    \vspace{-7pt}
    \caption{Latency decrease achieved by proposed MILP approach over extended version of HEFT under increasing TG size. Box plot in red shows overall distribution of latency decrease across all TG sizes.}
    \label{fig:syntheticLatency}
    \vspace{-5pt}
\end{figure}

The scalability of the proposed MILP method was evaluated experimentally using the 25 synthetic workflows we generated.
\cref{fig:syntheticLatency} demonstrates the latency improvement attained by our MILP approach over the enhanced version of HEFT, as the size of the workflows increased.
Notably, MILP achieved a significant latency decrease for all workflows of all TG sizes. 
Specifically, MILP provided a latency improvement over HEFT ranging from 16.56\% to 48.72\%, with a mean and median of 33.03\% and 33.6\%, respectively, across all TG sizes.
The latency decrease presented the highest variability for TGs with 10 nodes, due to the smaller number of tasks requiring specific capabilities.
This resulted in ETGs with more diverse structures, compared to the other cases.

\begin{figure}[!t]
    \centering
    \includegraphics[width=0.65\columnwidth]{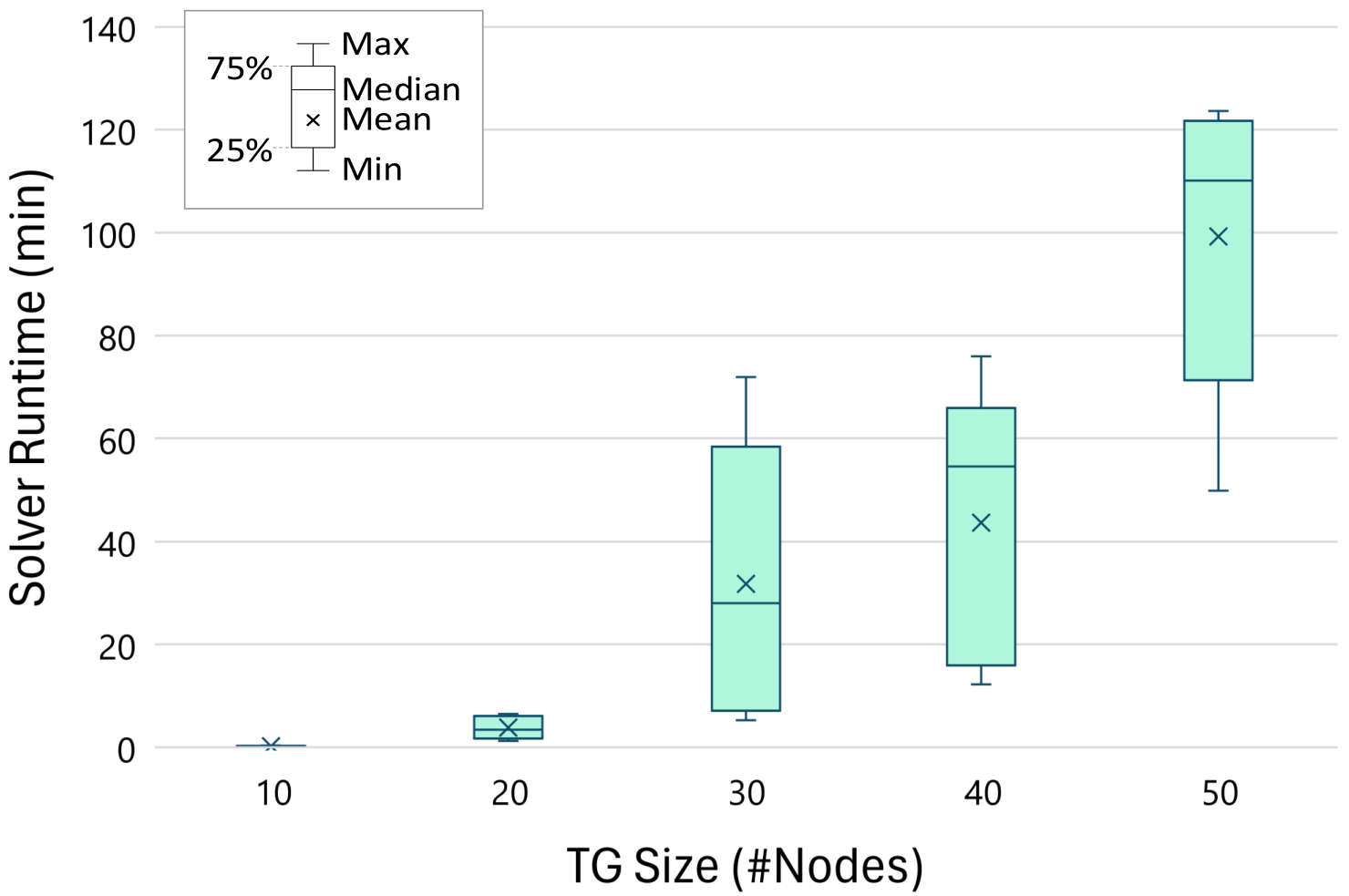}
    \vspace{-7pt}
    \caption{Solver runtime for proposed\,MILP\,approach under increasing TG size.}
    \label{fig:syntheticRuntime}
    \vspace{-5pt}
\end{figure}

\cref{fig:syntheticRuntime} shows the time required by the Gurobi solver to provide a solution for the proposed MILP technique as the TG size increased. The average solver execution time is also reported in \cref{tab:syntheticETGs}, along with the average number of variables/constraints in each case.
The solver runtime ranged from 3 to 20\,s for 10-node TGs, 1 to 6\,min for 20-node TGs, 5 to 72\,min for 30-node TGs, 12 to 76\,min for 40-node TGs, and 50 to 124\,min for 50-node TGs.
For reference, the runtime of the extended version of HEFT ranged from 71 to 4753\,ms.
Given the NP-hard nature of the examined problem \cite{Zengen2020}, the optimality of the proposed MILP approach in terms of latency, and considering that this is an offline scheduling method where the primary objective is the minimization of workflow execution time (rather than the solver runtime), the time required by Gurobi is reasonable, showcasing the scalability of our technique as the problem size increased.
The practicality of the proposed approach is further highlighted by the coarse-grained structure of relevant cyber-physical applications, which typically comprise 10--20 tasks, as demonstrated by the real-world workflow examined in \cref{subsec:real}. For these TG sizes, the solver provided the optimal solution in a short time frame, ranging from 3\,s to 6\,min.

\vspace{-3pt}
\section{Conclusion}
\label{sec:conclusion}
We proposed an offline approach to optimally schedule a workflow application in an edge-hub-cloud CPS with heterogeneous multicore processors and various sensing, actuating, or other specialized capabilities.
Our method utilizes a continuous-time MILP formulation to minimize the overall latency.
It comprehensively considers multiple constraints often ignored by existing scheduling approaches, both exact and heuristic. Specifically, it addresses the memory, storage, and energy limitations of the devices, the heterogeneity and multicore architecture of the processors, the distinct device capabilities, the execution deadline, as well as the computational and communication latency and energy requirements of the tasks.
Using a relevant real-world workflow, we compared our technique to the well-established HEFT scheduling heuristic, under different system configurations. In order for the comparison to be meaningful and fair, we extended HEFT by incorporating the deadline, capability, memory, storage, and energy constraints considered in our approach.
Furthermore, we investigated the scalability of our method using representative synthetic workflows of various sizes and appropriate parameters. 
The experimental results revealed that the proposed technique consistently outperformed HEFT, yielding an average latency decrease of 13.54\% in the real-world use case. 
Moreover, they demonstrated its scalability, as it provided an average latency improvement of 33.03\% over HEFT for the synthetic workflows, within a reasonable time frame.

\vspace{-3pt}
\section*{Acknowledgments}
\vspace{-1pt}
This work has been supported by the European Union’s Horizon 2020 research and innovation programme under grant agreement No. 739551 (KIOS CoE) and from the Government of the Republic of Cyprus through the Cyprus Deputy Ministry of Research, Innovation and Digital Policy.

\bibliographystyle{IEEEtran}

\vspace{-3pt}
\bibliography{references.bib}

\end{document}